\documentclass[paper]{JFM-FLM_Au}
\usepackage{amsmath,amssymb,graphicx,booktabs,float}
\usepackage[T1]{fontenc}
\usepackage[utf8]{inputenc}
\providecommand{\tightlist}{\setlength{\itemsep}{0pt}\setlength{\parskip}{0pt}}
\DeclareUnicodeCharacter{03B8}{\ensuremath{\theta}}
\DeclareUnicodeCharacter{03C0}{\ensuremath{\pi}}
\DeclareUnicodeCharacter{2248}{\ensuremath{\approx}}
\DeclareUnicodeCharacter{2082}{\ensuremath{_2}}
\DeclareUnicodeCharacter{2084}{\ensuremath{_4}}
\DeclareUnicodeCharacter{00B2}{\ensuremath{^2}}
\DeclareUnicodeCharacter{00B3}{\ensuremath{^3}}
\DeclareUnicodeCharacter{00B5}{\ensuremath{\mu}}
\DeclareUnicodeCharacter{03BC}{\ensuremath{\mu}}
\DeclareUnicodeCharacter{2009}{\,}
\DeclareUnicodeCharacter{202F}{\,}
\input{submission_cleanup}
\lefttitle{R. Cao, F. Luo, M. Bu, Y. Shi, H. Wen and B. Wang}
\righttitle{Wave structures in a non-premixed rotating detonation combustor}
\title{Three-dimensional wave structures and their formation mechanisms in a non-premixed hydrogen--air rotating detonation combustor}
\author{Ruiming Cao\aff{1}, Faxuan Luo\aff{1}, Mingyang Bu\aff{1},\\ Yingchen Shi\aff{1}, Haocheng Wen\aff{2} \and Bing Wang\aff{2}}
\affiliation{\aff{1}School of Aerospace Engineering, Tsinghua University, Beijing 100084, China
\aff{2}Institute for Aero Engine, Tsinghua University, Beijing 100084, China}
\corresau{Haocheng Wen, \email{wen@tsinghua.edu.cn}}
\hypersetup{hidelinks}
\begin{document}
\maketitle
\begin{abstract}
\emergencystretch=2em
Three-dimensional large-eddy simulations (LES) are performed for a non-premixed hydrogen--air rotating detonation combustor (RDC) and validated against experimental high-frequency pressure signals and OH* chemiluminescence features. Combined analysis of experimental snapshots and numerical results identifies the primary detonation wave (PDW), downstream oblique shock (DOS), detached detonation waves (DDWs), and inward-transmitted shock (ITS). Their three-dimensional structures and interactions are examined. Recirculation behind the backward-facing step (BFS) retains hot products near the inner wall, whereas fresh reactants occupy the outer-wall region. Discrete hydrogen jet columns (HJCs) are distributed within the fresh reactants. The PDW and its principal heat-release region propagate mainly through the outer fresh-mixture layer. Multiple DDWs successively separate from the upstream side of the PDW and propagate upstream along the HJCs. DDW evolution comprises detonation-front deformation, detonation-front separation, upstream propagation along the HJCs, heat-release decay and recovery near the injectors. DDW formation arises from competition between PDW-driven circumferential propagation and upstream propagation supported by the local fuel distribution. The DDWs generate diffracted shocks that influence the wave system. Pressure disturbances transmitted from the PDW into the hot-product layer form an ITS that leads the PDW in circumferential phase. The ITS exhibits reduced net heat release near its front and induces transmitted and reflected oblique shocks in the fresh-mixture layer. Periodic blast waves generated by DDW shock diffraction propagate inward and catch up with the ITS, producing pressure disturbances that may sustain its propagation. These findings connect experimental emission features to three-dimensional wave structures and clarify their formation and interaction.
\end{abstract}

\section{Introduction}\label{sec:1}

A rotating detonation combustor (RDC) continuously supplies reactants to sustain one or more detonation waves propagating azimuthally through the chamber. Rapid heat release and high power density make RDCs attractive for compact propulsion and power-generation devices \citep{wolanskiDetonativePropulsion2013,rankinOverviewPerformanceApplication2017}. Following initial ignition, continuous reactant supply sustains the rotating waves without repeated ignition \citep{anandRotatingDetonationCombustors2019}. RDCs also offer the potential for pressure-gain combustion \citep{rankinOverviewPerformanceApplication2017,ramanNonidealitiesRotatingDetonation2023}. However, nonideal phenomena in practical combustors can introduce losses and limit the realisation of theoretical performance gains. \citet{ramanNonidealitiesRotatingDetonation2023} reviewed these phenomena, including unsteady mixing, detonation-wave propagation in inhomogeneous mixtures, wave multiplicity and multiple competing waves, secondary combustion, and secondary waves. Their review describes how injector recovery controls fuel--air mixing, postdetonation-gas entrainment, and the development of a stratified, partially combusted predetonation mixture. Secondary waves also shape the fuel--oxidiser stratification and deflagrated regions encountered by the primary detonation waves. Understanding these coupled processes therefore requires examining the spatial structures, interactions, and formation mechanisms of the primary detonation and its accompanying waves. For non-premixed configurations, linking experimentally observed emission features to the corresponding three-dimensional waves and local reaction structures provides a basis for understanding their physical origins.

Two-dimensional simulations established the basic flow-field picture comprising the primary detonation, trailing oblique shock, contact discontinuities, and refill region \citep{hishidaFundamentalsRotatingDetonations2009,schwerNumericalInvestigationPhysics2011}. Subsequent analysis identified losses associated with combustion outside the detonation front and compression across accompanying shocks \citep{schwerNumericalInvestigationPhysics2011}. Studies then examined local reactions and wave-generation mechanisms. Using two- and three-dimensional simulations, \citet{uemuraTransverseWaveGeneration2013} showed that periodic compression and burning of unreacted pockets near the junction of the primary wave, oblique shock, and product interface generate micro-explosions and transverse waves. Reactant inhomogeneity can also cause velocity deficits. \citet{fujiiNumericalInvestigationDetonation2017} found that increasing the width of burned-gas layers reduced detonation propagation speed and increased the amount of incompletely burned fuel, with a stronger effect under non-premixed conditions. Locally lean mixtures can also increase induction times \citep{rajInfluenceFuelInhomogeneity2024}. Severe local fuel depletion can additionally cause shock--reaction decoupling \citep{rajInfluenceFuelInhomogeneity2024}. For discrete premixed injection, \citet{ullmanInfluenceInjectorConfiguration2024} found that shock--reaction recoupling as a wave entered a reactant jet produced local explosions, generating transverse waves and counter-propagating secondary waves. \citet{valenciaFlowfieldAnalysisPerformance2024} linked reverse compression waves to the alternating interaction of the detonation with fresh reactants and burned-gas regions between discrete inlets. These waves distorted the contact surface and perturbed the inlet mass flow. Complementary studies of detonations in idealised inhomogeneous media showed that periodic reactant-concentration variations can produce shock--reaction decoupling, local reinitiation, and double cellular structures \citep{wangNumericalStudiesDetonation2023}. Transverse temperature discontinuities modify propagation in hot and cold layers, generate oblique shocks connecting different front segments, and inhibit transverse-wave propagation at sufficiently large temperature differences \citep{maNumericalStudyPropagation2025}. Beyond axial--azimuthal unwrapped models, radial--azimuthal curved-channel simulations revealed curvature constraints on stable propagation \citep{nejaamtheenEffectsDimensionlessRadius2021}. These studies link local mixture variations and heat release to wave generation. Three-dimensional simulations are needed to resolve axial mixing, radial confinement, and the spatial organisation of discrete jets simultaneously in the actual combustor geometry.

Three-dimensional studies show how wall geometry influences the radial organisation of the wave system. For fixed chamber width, length, and injection conditions, \citet{zhouNumericalInvestigationShock2013} observed more pronounced near-head shock reflections as the inner radius increased. The outer-wall curvature also promotes continuous detonation propagation in hollow configurations \citep{tangThreedimensionalNumericalInvestigations2015}. Channel width also changes the relative positions and strengths of the inner- and outer-wall fronts \citep{kattaEffectIncreasingChannel2019}. \citet{kattaEffectIncreasingChannel2019} combined three-dimensional simulations and chemiluminescence imaging to show that increasing channel width advances the inner-wall front relative to the outer-wall front and increases the outer-wall peak pressure, while the inner-wall peak pressure remains nearly unchanged. The simulations and measurements also reveal compression associated with oblique shocks between the walls. Experiments in hollow configurations further revealed radial stratification of the coupled reaction zone and inward-diffracting shocks \citep{huangAnalysisRadialStructure2023}, while propagation along a wall-detached reactant layer induces oblique shocks on both sides \citep{huangPropagationFlowField2023}. These results show how wall confinement and reactant-layer location influence wave organisation, motivating identification of primary waves, nonreactive shocks, and local reaction zones through their propagation media and spatial relationships.

Non-premixed three-dimensional studies show that wave-induced pressure changes and injector response jointly control refill-mixture formation \citep{sunInteractionRotatingDetonation2019,satoMixingDetonationStructure2021a,nassiniCharacterizationRefillRegion2023}. Pressure rises modify residence and mixing times \citep{sunInteractionRotatingDetonation2019}, while differences in fuel and oxidiser blockage and recovery produce local composition variations \citep{satoMixingDetonationStructure2021a,nassiniCharacterizationRefillRegion2023} and contribute to axial and radial stratification \citep{nassiniCharacterizationRefillRegion2023}. Using large-eddy simulations (LES) of the complete injection system, \citet{nassiniCharacterizationRefillRegion2023} characterised axial and radial stratification in the refill region. Their analysis suggested that transient local wave-speed variations can occur after the wave encounters pure-air pockets; this behaviour was stochastic and did not occur at the injector-passage frequency. Mixing state affects the partition of fuel consumption between detonation and deflagration regimes \citep{strempflEffectsMixingAssumptions2024}, whereas upstream temperature and composition inhomogeneities modify three-dimensional front morphology \citep{bonanniCouplingDetonationStructure2024a}. Secondary waves also contribute to injector perturbations \citep{nassiniCharacterizationRefillRegion2023,vignatAnalysisWeakSecondary2024}. High-pressure regions formed during primary--secondary wave interactions alter injector blockage, recovery, and local heat release \citep{vignatAnalysisWeakSecondary2024}. In disk-shaped rocket configurations, discrete and slit injection produce different near-wall reactant structures and product-residence characteristics \citep{nejaamtheenThreedimensionalMixingDetonation2026}. In addition to injection-induced inhomogeneity, recirculation behind a backward-facing step retains hot products and alters the leading-wave morphology \citep{athmanathanEffectsReactantStratification2022}. Related simulations under high-flight-Mach-number conditions also linked recirculation-induced temperature stratification to leading wall-attached shocks \citep{chenEffectsBackwardfacingStep2026}. Combining three-dimensional unsteady Reynolds-averaged Navier--Stokes (URANS) simulations and high-speed imaging, \citet{athmanathanEffectsReactantStratification2022} showed that stratification allows reactants to remain unburned after passage of the primary detonation and subsequently burn in a wall-curvature-induced azimuthal reflected-shock combustion (ARSC) region. Together, these studies connect injector response and reactant stratification to local reaction structures and wave-front variations.

Advances in experimental diagnostics have made these coupled processes increasingly observable. Jet-mixing measurements in nonreacting water models revealed jet trajectories and entrainment \citep{bluemnerExperimentalStudyReactant2019}, while acetone planar laser-induced fluorescence (PLIF) in reacting flows identified post-wave injection suppression and recovery \citep{rankinEvaluationMixingProcesses2016}. In the partially premixed hydrogen--air configuration examined by \citet{burkeEffectPremixedStratification2021}, improved local mixing was associated with higher wave speeds. High-speed chemiluminescence revealed lifted fronts in ethylene--air experiments \citep{choHighRepetitionRateChemiluminescenceImaging2016}; mid-infrared imaging identified accompanying oblique and exit-reflected shocks in a non-premixed hydrogen--air RDC \citep{rankinInvestigationStructureDetonation2019}; and comparisons of emission and pressure distributions located concentrated reaction near the head of a cylindrical combustor \citep{yokooExperimentalStudyInternal2021}. Simultaneous 500-kHz OH-PLIF and OH* chemiluminescence further identified deflagrative preburning during refill \citep{wang500kHzOHPLIF2024}. \citet{daiHighPerformanceUltravioletOptical2026} developed a high-gain ultraviolet optical sensing system that uses synchronised and interleaved triggering of multiple intensified complementary metal--oxide--semiconductor (ICMOS) cameras to obtain multiband chemiluminescence measurements and short image sequences with high temporal resolution. Synchronised end-view and side-view imaging enabled reconstruction of primary-wave, trailing-shock, and refill-layer features \citep{armbrusterVisualizationDetonationWave2026}. Pressure--emission analysis provided evidence of secondary-wave interactions \citep{chaconSecondaryWavesDynamics2021}, while time-resolved measurements of temperature and water-vapour concentration revealed the influence of combustion products on the local mixture state \citep{fuggerDynamicsNonpremixedRotating2021}. These diagnostics reveal spatial and temporal signatures of the flow; comparisons with resolved numerical fields help distinguish the waves and reaction zones contributing to each emission feature.

Direct comparisons between experimental observations and numerical results have advanced the identification of visible structures. \citet{cocksHighFidelitySimulations2016} compared mean experimental OH* images with instantaneous computed heat-release fields to assess the predicted detonation-wave height. Their three-dimensional simulations also showed inner-wall flameholding and an oblique shock leading the detonation front. Comparisons of synthetic and experimental OH-PLIF also identified contact burning and near-injector parasitic combustion \citep{satoMixingDetonationStructure2021a}. \citet{hellardQuasiCJRotatingDetonation2024} evaluated OH* concentrations in LES postprocessing using the quasi-steady-state approximation (QSSA) and qualitatively compared the predicted emission structures with experimental chemiluminescence images. The simulation also reproduced the detonation propagation velocity, wave structure, and burnt-gas expansion in a partially premixed methane--oxygen combustor. These studies demonstrate how imaging and three-dimensional calculations can connect experimental signatures to specific waves and reaction zones \citep{kattaEffectIncreasingChannel2019,athmanathanEffectsReactantStratification2022,cocksHighFidelitySimulations2016,hellardQuasiCJRotatingDetonation2024}.

The present study combines high-fidelity three-dimensional large-eddy simulations with experimental high-frequency pressure measurements and OH* chemiluminescence snapshots to examine the wave system in a non-premixed hydrogen--air RDC. The numerical results are compared with the measured pressure and emission features before the resolved species, heat-release, and shock distributions are used to identify the corresponding flow structures. The wave system comprises the primary detonation wave (PDW), a downstream oblique shock (DOS) induced by the PDW, several detached detonation waves (DDWs) induced by the PDW and propagating upstream along discrete hydrogen jet columns (HJCs), and an inward-transmitted shock (ITS) formed by pressure disturbances transmitted from the PDW into the hot-product layer. Particular attention is paid to the spatial position and propagation characteristics of the PDW, its interactions with discrete HJCs, and the formation and evolution of DDWs and the ITS. Time-resolved flow fields are then analysed to examine interactions between the ITS and pressure disturbances generated during DDW evolution.

\newpage
The remainder of this paper is organised as follows. §\,\ref{sec:2} describes the combustor configuration, numerical methods, and computational settings. §\,\ref{sec:3.1} assesses the numerical results against measured pressure-cycle periods, waveforms, and chemiluminescence features, while §\,\ref{sec:3.2} presents the global flow-field structures. §\,\ref{sec:4.1} examines the PDW and its interactions with the HJCs. §\,\ref{sec:4.2} investigates DDW formation and propagation, and §\,\ref{sec:4.3} discusses ITS formation and its interactions with the post-wave flow field and blast waves. Finally, §\,\ref{sec:5} summarises the main conclusions.

\begin{figure}[!t]
\centering
\includegraphics[width=\textwidth,height=0.77\textheight,keepaspectratio]{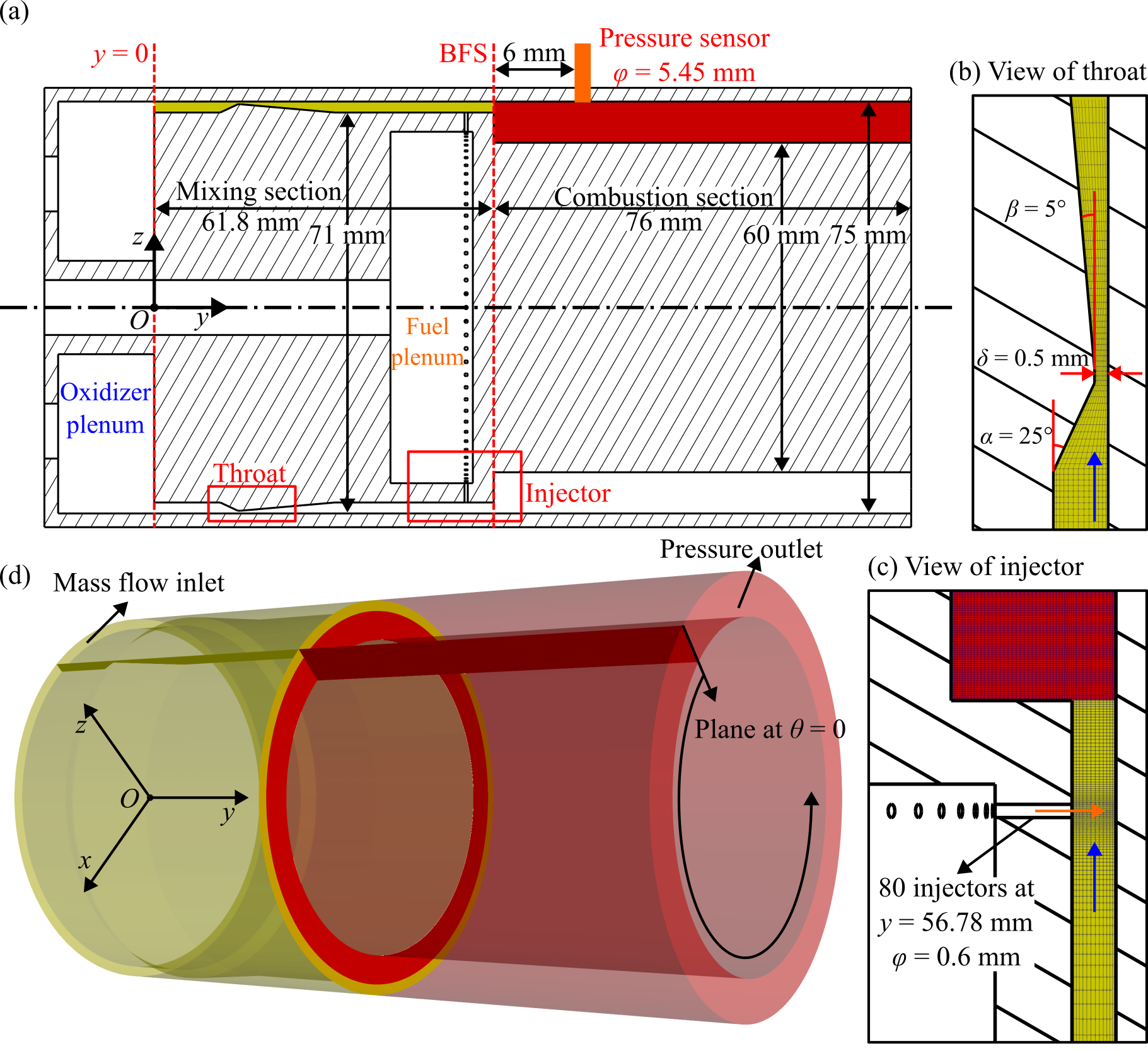}
\caption{Combustor configuration, local meshes, and computational domain: (a) axial cross-section and principal dimensions; (b) enlarged view of the mixing-section throat and its mesh; (c) enlarged view of the fuel-injection region and its mesh; and (d) three-dimensional computational domain, coordinate systems, and inlet/outlet boundary locations.}\label{fig:1}
\end{figure}

\section{Methodology}\label{sec:2}

\subsection{Physical model}\label{sec:2.1}

\subsubsection{Experimental configuration}\label{sec:2.1.1}

The present study considers a non-premixed hydrogen--air rotating detonation combustor with axial oxidiser supply and radial fuel injection. The combustor configuration and injection arrangement are based on those used in previous experimental investigations \citep{daiHighPerformanceUltravioletOptical2026,shiCalculationModelWall2025,buRotatingDetonationBladeless2026}. The combustor consists of an upstream mixing section and a downstream combustion section connected by a backward-facing step (BFS), as shown in figure~\ref{fig:1}(a). The axial coordinate originates at the mixing-section inlet, \(y = 0\,\mathrm{mm}\), and the BFS is located at \(y = 61.8\,\mathrm{mm}\).

The mixing section comprises an annular converging--diverging passage with inner and outer diameters of 71 and \(75\,\mathrm{mm}\), respectively, and an axial length of \(61.8\,\mathrm{mm}\). The combustion section is a constant-area annular passage with inner and outer diameters of 60 and \(75\,\mathrm{mm}\), respectively, and an axial length of \(76\,\mathrm{mm}\). Figure~\ref{fig:1}(b) shows the mixing-section throat, which has an upstream contraction angle of 25°, a downstream expansion angle of 5°, and an annular gap of \(0.5\,\mathrm{mm}\).

After passing through the plenum, the oxidiser enters the mixing section axially. Hydrogen is injected radially through 80 circular orifices uniformly distributed around the inner circumference of the mixing section. Each orifice has a diameter of \(0.6\,\mathrm{mm}\) and is centred at \(y = 56.78\,\mathrm{mm}\), as shown in figure~\ref{fig:1}(c). The radial hydrogen jets intersect the axial oxidiser crossflow, and the resulting mixture subsequently enters the downstream combustion section.

\subsubsection{Computational domain and mesh}\label{sec:2.1.2}

The computational domain includes the mixing and combustion sections. The three-dimensional domain and coordinate-system definitions are shown in figure~\ref{fig:1}(d). A multiblock body-fitted mesh is used to resolve the annular passage and geometric variations near the throat and BFS. The \(y\) axis of the global Cartesian coordinate system \((x,y,z)\) coincides with the combustor axis, and the \(x\)-\(z\) plane at \(y = 0\,\mathrm{mm}\) corresponds to the mixing-section inlet. A coaxial cylindrical coordinate system \((\theta,r,y)\) is introduced to analyse circumferential wave propagation and radial flow organisation. The zero-angle half-plane is defined by \(x = 0\) and \(z > 0\), and \(\theta\) increases counterclockwise when viewed axially from the combustion-section exit. The coordinate transformation is given by (\ref{eq:1}).

\begin{equation}
r = \sqrt{x^{2} + z^{2}},\quad\quad\theta = \operatorname{mod}\left\lbrack \operatorname{atan2}(x,z),\, 2\pi \right\rbrack.
\label{eq:1}\end{equation}

The combustion and mixing sections contain \(2000 \times 49 \times 350\) and \(2000 \times 13 \times 200\) cells, respectively, in the \(\theta\), \(r\), and \(y\) directions. The domain therefore contains approximately \(3.95 \times 10^{7}\) cells, with an average cell size of approximately \(0.15\,\mathrm{mm}\). Figure~\ref{fig:1}(b) shows the locally refined mesh near the mixing-section throat. Figure~\ref{fig:1}(c) shows the mesh near the BFS and fuel-injection region. The average cell sizes are approximately \(0.1\,\mathrm{mm}\) in the detonation propagation region and \(0.05\,\mathrm{mm}\) near the fuel-injection orifices.

\subsection{Experimental setup}\label{sec:2.2}

The experiments were conducted using the experimental facility described in previous studies \citep{daiHighPerformanceUltravioletOptical2026,shiCalculationModelWall2025,buRotatingDetonationBladeless2026}. Details of the experimental system, triggering sequence, and operating procedures are provided in these references. The high-frequency wall-pressure signals and OH* chemiluminescence snapshots obtained from this facility are used to assess the numerical results.

Wall-pressure signals are acquired at a sampling frequency of \(800\,\mathrm{kHz}\). The pressure transducer is mounted on the outer wall, with a sensing-face diameter of \(5.45\,\mathrm{mm}\) and its centre located \(6\,\mathrm{mm}\) downstream of the BFS, as indicated by the orange region in figure~\ref{fig:1}(a). For comparison, the numerical pressure signal is extracted at the corresponding wall location and sampled at the experimental sampling frequency. An area-weighted average of the wall pressure is applied over the corresponding sensing-face region.

OH* chemiluminescence snapshots are obtained using the high-performance ultraviolet optical sensing system developed by \citet{daiHighPerformanceUltravioletOptical2026}. A narrowband filter centred at \(310\,\mathrm{nm}\) is mounted in front of the camera lens to select the OH* emission band and suppress out-of-band background signals. The image resolution is \(1608 \times 1104\) pixels, and the exposure time is \(200\,\mathrm{ns}\) to balance the signal-to-noise ratio against motion blur associated with rapidly propagating waves.

\subsection{Numerical setup}\label{sec:2.3}

\subsubsection{Governing equations}\label{sec:2.3.1}

The numerical model solves the three-dimensional, unsteady, compressible multicomponent Navier--Stokes equations with chemical source terms. A body-fitted coordinate transformation maps the physical coordinates \((x,y,z)\) onto the computational coordinates \((\xi,\eta,\zeta)\). The governing equations are expressed in conservative form as (\ref{eq:2}).

\begin{equation}
\frac{\partial(\mathcal{J}\boldsymbol{Q})}{\partial t} + \frac{\partial{\widehat{\boldsymbol{F}}}_{\xi}}{\partial\xi} + \frac{\partial{\widehat{\boldsymbol{F}}}_{\eta}}{\partial\eta} + \frac{\partial{\widehat{\boldsymbol{F}}}_{\zeta}}{\partial\zeta} = \frac{\partial{\widehat{\boldsymbol{F}}}_{\xi,v}}{\partial\xi} + \frac{\partial{\widehat{\boldsymbol{F}}}_{\eta,v}}{\partial\eta} + \frac{\partial{\widehat{\boldsymbol{F}}}_{\zeta,v}}{\partial\zeta}+\mathcal{J}\boldsymbol{S}.
\label{eq:2}\end{equation}

Here, \(\mathcal{J} = \det\lbrack\partial(x,y,z)/\partial(\xi,\eta,\zeta)\rbrack\) is the Jacobian determinant of the coordinate transformation. The vectors \(\boldsymbol{Q}\) and \(\boldsymbol{S}\) denote the conserved variables and chemical source terms, respectively, while \({\widehat{\boldsymbol{F}}}_{\alpha}\) and \({\widehat{\boldsymbol{F}}}_{\alpha,v}\) are the inviscid and viscous fluxes in computational direction \(\alpha\). The conserved variables and source terms are defined as (\ref{eq:3}).

\begin{equation}
\begin{aligned}
\boldsymbol{Q} & = \left\lbrack \rho,\rho u,\rho v,\rho w,\rho E,\rho Y_{1},\ldots,\rho Y_{N_{s} - 1} \right\rbrack^{T}, \\
\boldsymbol{S} & = \left\lbrack 0,0,0,0,0,{\dot{\omega}}_{1},\ldots,{\dot{\omega}}_{N_{s} - 1} \right\rbrack^{T}.
\end{aligned}
\label{eq:3}\end{equation}

For compact notation, let \((x_{1},x_{2},x_{3}) = (x,y,z)\) and \((u_{1},u_{2},u_{3}) = (u,v,w)\), with \(j = 1,2,3\) indexing the Cartesian directions. The computational-space fluxes are obtained from the physical-space fluxes through:

\begin{equation}
{\widehat{\boldsymbol{F}}}_{\alpha}=\mathcal{J}\sum_{j = 1}^{3}\frac{\partial\alpha}{\partial x_{j}}\boldsymbol{F}_{j},\quad\quad{\widehat{\boldsymbol{F}}}_{\alpha,v}=\mathcal{J}\sum_{j = 1}^{3}\frac{\partial\alpha}{\partial x_{j}}\boldsymbol{F}_{j,v},\quad\quad\alpha \in \{\xi,\eta,\zeta\}.
\label{eq:4}\end{equation}

\begin{equation}
\boldsymbol{F}_{j} = \begin{bmatrix}
\rho u_{j} \\
\rho uu_{j} + p\delta_{1j} \\
\rho vu_{j} + p\delta_{2j} \\
\rho wu_{j} + p\delta_{3j} \\
(\rho E + p)u_{j} \\
\rho Y_{1}u_{j} \\
 \vdots \\
\rho Y_{N_{s} - 1}u_{j}
\end{bmatrix},\quad\quad\boldsymbol{F}_{j,v} = \begin{bmatrix}
0 \\
\tau_{1j} \\
\tau_{2j} \\
\tau_{3j} \\
\sum_{k = 1}^{3}u_{k}\tau_{kj} - q_{j} - \sum_{i = 1}^{N_{s} - 1}h_{i}j_{i,j} \\
 - j_{1,j} \\
 \vdots \\
 - j_{N_{s} - 1,j}
\end{bmatrix}.
\label{eq:5}\end{equation}

The quantities \(\rho\), \(p\), and \(E\) denote density, pressure, and total energy per unit mass, respectively. For species \(i\), \(Y_{i}\), \(h_{i}\), \(j_{i,j}\), and \({\dot{\omega}}_{i}\) are the mass fraction, specific enthalpy, mass-diffusion flux in the \(x_{j}\) direction, and net mass production rate, respectively. The number of species is \(N_{s}\), and \(\delta_{ij}\) is the Kronecker delta.

The gas mixture is described using the ideal-gas equation of state.

Hydrogen/air combustion is modelled using the chemical kinetic mechanism of \citet{hishidaNumericalSimulationPulsed1993}, which contains 9 species and 19 elementary reactions. Transport equations are explicitly solved for eight reactive species: H₂, O₂, OH, H, O, H₂O, HO₂, and H₂O₂. The mass fraction of the inert species N₂ is determined from the mass-fraction constraint. The mechanism extension used to predict OH* emission is described in §\,\ref{sec:2.3.4}.

\subsubsection{Boundary and initial conditions}\label{sec:2.3.2}

No-slip, adiabatic boundary conditions are imposed on the inner wall, outer wall, and BFS wall. The mixing-section inlet is prescribed as a constant-mass-flow boundary with an oxidiser flow rate of \(130\,\mathrm{g\,s^{-1}}\) and a static temperature of \(298\,\mathrm{K}\). The oxidiser contains \(40\,\%\) oxygen by volume, corresponding to O₂ and N₂ mass fractions of 0.4324 and 0.5676, respectively. A no-backflow pressure-outlet boundary is applied at the combustion-section exit. The oxidiser-inlet and pressure-outlet locations are indicated in figure~\ref{fig:1}(d).

Hydrogen injection is represented by a boundary model based on the approach described by \citet{fievisohnSteadyStateAnalysisRotating2017}. The model combines isentropic relations within the nozzle with conservation of mass, momentum, and energy across the transition region. It determines the injection mass flow rate and exit state from the total pressure, total temperature, area ratio, and downstream flow state, accounting for both choked and unchoked injection. The hydrogen total pressure and total temperature are \(725\,\mathrm{kPa}\) and \(300\,\mathrm{K}\), respectively. Pure hydrogen is supplied with an area ratio of 0.5. Ghost cells transfer conserved variables between the mixing- and combustion-section grid blocks.

The simulation is initialised by advancing the cold flow with chemical reactions disabled until a developed hydrogen distribution is established. Reactions are then activated, and an ignition region is imposed near the BFS in the combustion section. The region occupies one-eighth of the circumference, spans the full radial width of the passage, and extends \(5\,\mathrm{mm}\) downstream of the BFS. The ignition pressure and temperature are \(20\,\mathrm{MPa}\) and \(3000\,\mathrm{K}\), respectively. After the detonation has propagated stably for ten rotation cycles, subsequent flow fields are extracted for pressure, emission, and wave-structure analyses.

\subsubsection{Solver configuration}\label{sec:2.3.3}

The simulations employ a three-dimensional version of JANC, a JAX-based compressible reactive-flow solver. The underlying JANC framework is described in \citet{wenJANCCosteffectiveDifferentiable2026}, while the present calculations perform LES of three-dimensional reacting flow on a multiblock body-fitted mesh. The grid metrics satisfy the geometric conservation law to maintain numerical conservation after coordinate transformation.

The governing equations are discretised using the finite-volume method. Convective fluxes are evaluated using a Godunov-type method, with interface conservative variables reconstructed by the monotonic upstream-centred scheme for conservation laws (MUSCL) and interface Riemann problems solved using the Harten--Lax--van Leer--contact (HLLC) approximate Riemann solver. Viscous fluxes are discretised using a sixth-order central-difference scheme. The flow terms are advanced using a third-order total-variation-diminishing (TVD) Runge--Kutta method. Chemical source terms are treated through operator splitting and a point-implicit method to address the stiffness associated with chemical timescales.

Thermodynamic properties are evaluated using seven-coefficient National Aeronautics and Space Administration (NASA) polynomials, and dynamic viscosity is determined using Sutherland's law. The molecular Prandtl and Schmidt numbers are both set to 0.7. Turbulence is represented using LES with the Vreman subgrid-scale model, and the turbulent Prandtl and Schmidt numbers are both set to 0.72. The calculations are performed in parallel on eight Tesla V100-SXM2-32GB graphics processing units (GPUs).

\subsubsection{Numerical visualisation method}\label{sec:2.3.4}

To connect the experimental OH* chemiluminescence features with the numerical reaction structures, the base hydrogen--oxygen mechanism is extended to include OH* formation, radiative decay, and collisional quenching \citep{zhaoExperimentalNumericalStudy2018,petersenCalibrationChemicalKinetics2003,smithLowPressureFlame2002,tamuraCollisionalQuenchingCHA1998}. The corresponding reaction forms are listed in (\ref{eq:6}), where M denotes a third-body collision partner and \(h\nu\) is the energy of the emitted photon.

\begin{equation}
\begin{aligned}
H + O + M & \rightleftharpoons OH^{*} + M, \\
OH^{*} & \rightarrow OH + h\nu, \\
OH^{*} + M & \rightleftharpoons OH + M.
\end{aligned}
\label{eq:6}\end{equation}

For a specified emission transition, the emitted signal depends on the excited-state number density, radiative transition probability, and optical collection factors. \citet{giassiAnalysisCHConcentration2016} used this relationship for CH* emission. For the qualitative OH* visualisation used here, the local emission is modelled as proportional to the OH* number density, and hence its mass density, as expressed in (\ref{eq:7}).

\begin{equation}
\varepsilon_{OH^{*}} \propto n_{OH^{*}} \propto \rho_{OH^{*}}.
\label{eq:7}\end{equation}

Under an optically thin approximation, with absorption along the optical path neglected, the camera signal is formed by the accumulation of local emission along the viewing direction. The general line-of-sight expression is therefore given by (\ref{eq:8}), where \(\mathcal{L}\) denotes a viewing ray and \(s\) is the path coordinate along that ray.

\begin{equation}
I_{\mathrm{cam}} \propto \int_{\mathcal{L}}^{}\varepsilon_{OH^{*}}\, ds \propto \int_{\mathcal{L}}^{}\rho_{OH^{*}}\, ds.
\label{eq:8}\end{equation}

In the present postprocessing, radial integration is used to construct the OH* surface-density distribution in the \((\theta,y)\) plane, as expressed in (\ref{eq:9}). The quantities \(R_{1}(y)\) and \(R_{2}(y)\) are the inner- and outer-wall radii at the corresponding axial location, and \(\sigma_{OH^{*}}\) is the OH* surface density, with units of \(\mathrm{kg\,m^{-2}}\). This projection is used for qualitative comparisons of reaction-zone locations and emission morphology with the experimental snapshots. A linear scale displays the principal emission structures, while a logarithmic scale enhances the visibility of weaker features.

\begin{equation}
\sigma_{OH^{*}}(\theta,y) = \int_{R_{1}(y)}^{R_{2}(y)}\rho_{OH^{*}}(\theta,r,y)\, dr.
\label{eq:9}\end{equation}

\section{Wave structures and propagation characteristics}\label{sec:3}

\subsection{Experimental evidence and validation}\label{sec:3.1}

A representative single-wave rotating detonation condition is selected, and the experimental and numerical operating parameters are listed in table~\ref{tab:1}. The oxidiser is oxygen-enriched air containing \(40\,\%\) oxygen by volume. The experimental oxidiser flow rate is \(127 \pm 5\,\mathrm{g\,s^{-1}}\), and the equivalence ratio is \(0.67 \pm 0.05\). The flow rates and equivalence ratios reported in the table are time-averaged values.

\begin{table}\centering
\begin{tabular}{lcc}
\hline
 & Experimental & Numerical \\
\hline
Oxidiser flow rate & $127\pm5\,\mathrm{g\,s^{-1}}$ & $130\,\mathrm{g\,s^{-1}}$ \\
Oxygen volume fraction & $40\,\%$ & $40\,\%$ \\
Equivalence ratio & $0.67\pm0.05$ & $0.66$ \\
\hline
\end{tabular}
\caption{Operating conditions for the experiment and simulation.}\label{tab:1}
\end{table}

Based on the representative operating condition listed in table~\ref{tab:1}, the high-frequency wall-pressure signals obtained from the experiment and numerical simulation were first compared to assess the ability of the numerical model to predict the propagation frequency and pressure-fluctuation characteristics of the rotating detonation wave.  

\begin{figure}
\centering
\includegraphics[width=\textwidth,height=0.77\textheight,keepaspectratio]{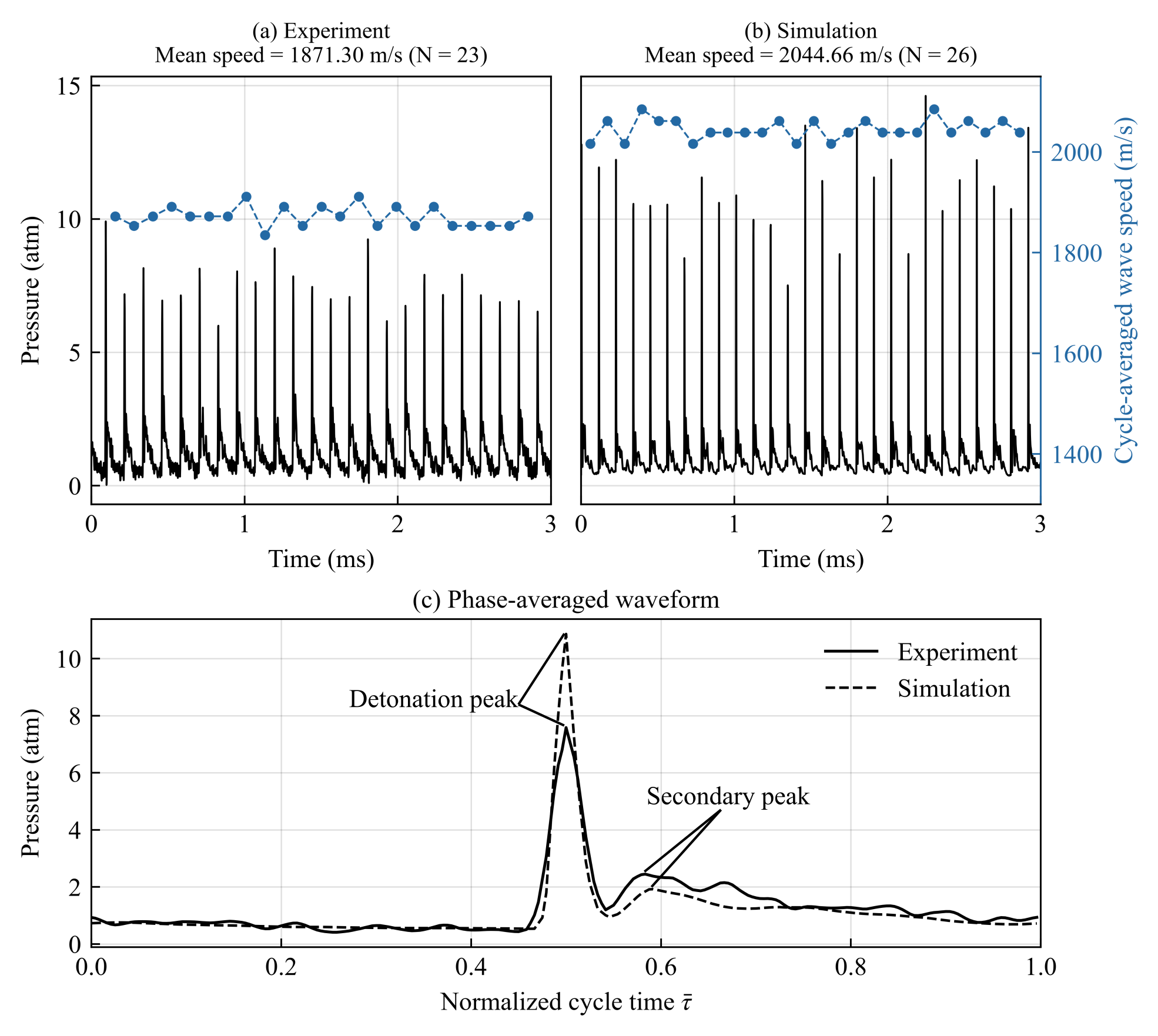}
\caption{Comparison of experimental and simulated pressure and wave-speed characteristics: (a) experimental pressure signal and cycle-averaged wave speed; (b) simulated pressure signal and cycle-averaged wave speed; and (c) phase-averaged pressure waveforms. In (a) and (b), black solid lines represent pressure signals, while blue circles connected by dashed lines represent cycle-averaged wave speeds plotted at the midpoint of each cycle.}\label{fig:2}
\end{figure}

Figure~\ref{fig:2} compares the experimental and numerical pressure signals and corresponding wave speeds over a 3-ms sampling window after the flow field reached a quasi-steady state. Both pressure signals exhibit clear, nearly periodic fluctuations. Cycle periods were determined from the intervals between successive primary pressure peaks, yielding 23 experimental and 26 numerical cycles. The experimental and numerical mean periods are 122.55 and \(112.16\,\mu\mathrm{s}\), respectively, with standard deviations of 1.33 and \(1.09\,\mu\mathrm{s}\). The corresponding coefficients of variation are \(1.09\,\%\) and \(0.97\,\%\), indicating small cycle-to-cycle variations within the selected window. The numerical mean period is approximately \(8.48\,\%\) shorter than the experimental value.

For the present single-wave mode, the cycle-averaged circumferential speed was calculated as \(D_i = 2\pi r/T_i\), where \(T_i\) is the interval between successive primary pressure peaks and \(r = 36.5\,\mathrm{mm}\) is the reference radius. These speeds are plotted at the midpoint of each cycle in figure~\ref{fig:2}(a) and figure~\ref{fig:2}(b). The overall mean speed was defined as \(\bar{D} = 2\pi r/\bar{T}\), equivalent to the total circumferential distance travelled over the analysed complete cycles divided by their total duration. The resulting experimental and numerical mean speeds were 1871.30 and \(2044.66\,\mathrm{m\,s^{-1}}\), respectively. The numerical prediction is approximately \(9.26\,\%\) higher than the experimental value. Both sets of cycle-averaged speeds fluctuate slightly around their respective means, consistent with the propagation stability indicated by the period statistics.

For phase averaging, each complete cycle was mapped onto a normalised cycle time, with its primary pressure maximum aligned at 0.5. As shown in figure~\ref{fig:2}(c), both waveforms contain a primary detonation peak followed by a secondary pressure peak. In the present simulation, the secondary pressure peak is associated with the trailing reflected-shock system. In discussing the high-speed wall-pressure measurements of \citet[figure~4(c)]{wenPropagationBehaviorsRotating2019}, \citet{athmanathanEffectsReactantStratification2022} noted a second weak pressure spike in the curved section that was absent in the linear section. Related shock-reflection and azimuthal reflected-shock combustion structures have also been investigated \citep{zhouNumericalInvestigationShock2013,kattaEffectIncreasingChannel2019,athmanathanEffectsReactantStratification2022}. The similar sequence and overall shape of these features indicate that the simulation reproduces the principal characteristics of the experimental pressure waveform.

Although the experimental and numerical signals were matched in terms of measurement location, sampling frequency, and spatial measurement range, the numerical pressure fluctuations remain noticeably larger. This difference may reflect the dynamic response of the pressure transducer, its mounting arrangement, signal transmission, and differences between the actual sensing-face response and the spatial averaging used in the simulation.

Overall, the simulation captures the nearly periodic pressure fluctuations, small cycle-to-cycle variations in wave speed, and double-peaked pressure waveform under the present operating condition, while predicting a shorter mean period, a higher mean wave speed, and larger pressure-fluctuation amplitudes. These comparisons support the subsequent use of the numerical flow field to investigate wave structures and their evolution.

\begin{figure}
\centering
\includegraphics[width=\textwidth,height=0.77\textheight,keepaspectratio]{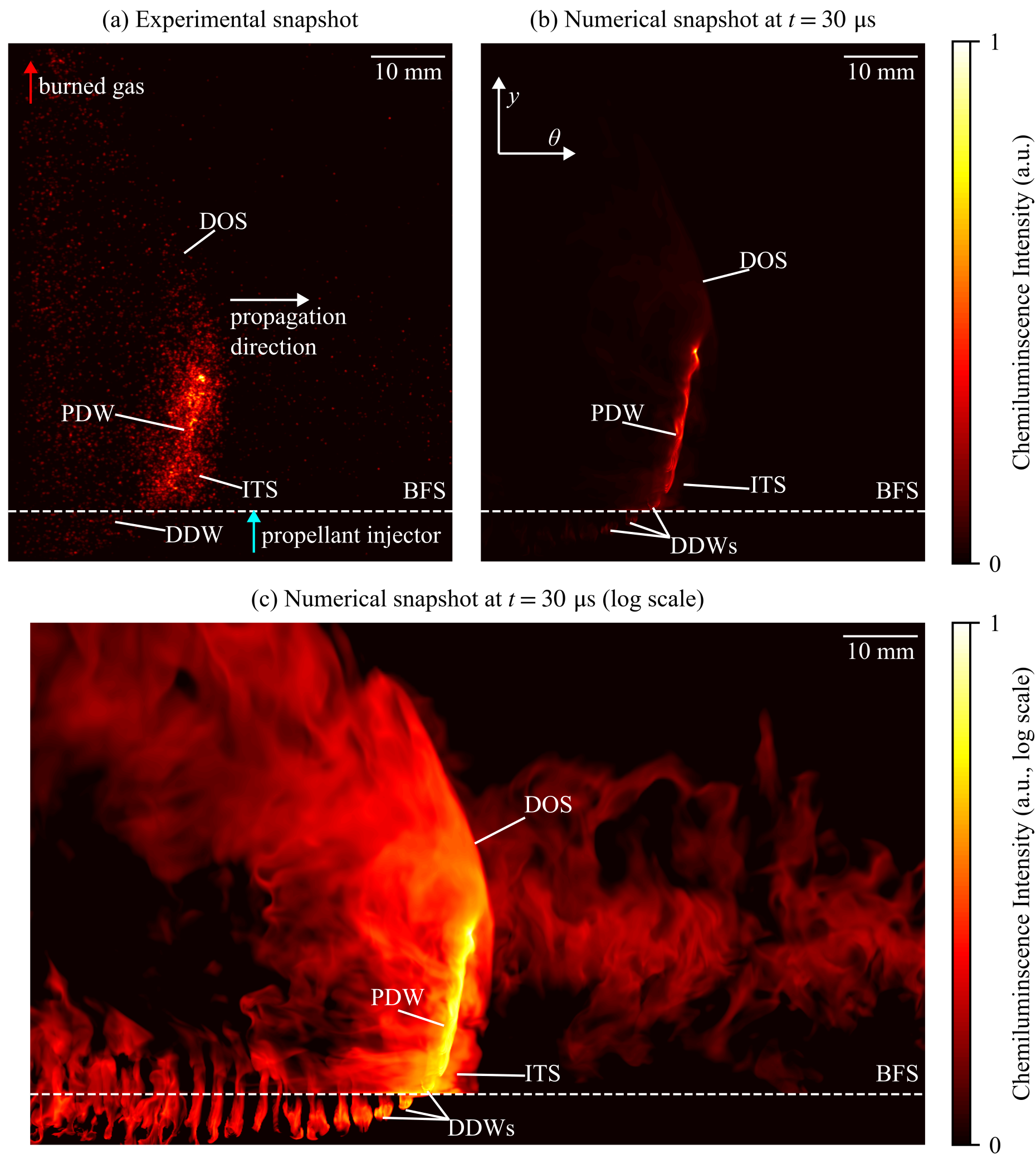}
\caption{Experimental and numerical OH* chemiluminescence distributions: (a) experimental snapshot, (b) numerical snapshot at \(t = 30\,\mu\mathrm{s}\) on a linear scale, and (c) numerical snapshot at \(t = 30\,\mu\mathrm{s}\) on a logarithmic scale.}\label{fig:3}
\end{figure}

Figure~\ref{fig:3}(a) and figure~\ref{fig:3}(b) show the experimentally measured OH* chemiluminescence snapshot and the instantaneous OH* surface density distribution obtained numerically at \(t = 30\,\mu\mathrm{s}\), respectively. Both the experimental and simulated results clearly capture the PDW, the DOS, and the ITS. In addition, both results exhibit faint emission features in the region associated with the DDWs identified in the resolved flow field. These features are difficult to distinguish clearly in the OH* snapshots alone.

Figure~\ref{fig:3}(c) shows the distribution in figure~\ref{fig:3}(b) on a logarithmic scale to reveal weaker emission features. Figure~\ref{fig:3}(c) clearly shows that the ITS remains phase-advanced relative to the PDW in the direction of propagation. It extends upstream towards the BFS and ultimately terminates at the BFS, exhibiting a propagation pattern attached near the BFS wall. The DDWs are located behind the PDW, continuously separate from its upstream side, and propagate upstream. The basis for naming these two structures and their formation mechanisms will be discussed in detail later.

Several experimental studies have shown that this distinctive DDW structure can be observed through chemiluminescence imaging in combustors with similar configurations (\citealp[figure~4]{wang500kHzOHPLIF2024}; \citealp[figure~4]{wangHighspeedChemiluminscenceOHPLIF2024}). However, the observability of detached detonation waves depends on the fuel and its chemical reaction characteristics. In the relevant study \citep[figure~4]{wang500kHzOHPLIF2024}, the chemiluminescence structure of the DDW was more distinct when methane was used as the fuel. When hydrogen was used, however, the emission signal was substantially weaker, making the wave difficult to identify directly. The DDWs cannot be clearly resolved in the OH* chemiluminescence snapshots obtained experimentally or numerically in the present study, possibly because the corresponding OH* emission intensity is relatively low.

Overall, the numerical results reproduce the main pressure-waveform and OH* emission features, supporting the subsequent analysis of the three-dimensional wave system.

\subsection{Three-dimensional wave structures}\label{sec:3.2}

In an RDC, the combustor geometry and injection system jointly govern the replenishment, mixing, and reaction of the fresh mixture, thereby determining the propagation mode of the detonation wave and the spatial distribution of its associated wave system \citep{anandRotatingDetonationCombustors2019,ramanNonidealitiesRotatingDetonation2023,zhouNumericalInvestigationShock2013,ullmanInfluenceInjectorConfiguration2024}. To establish the overall physical picture needed for the subsequent analysis of local structures, this section first examines the characteristic global flow-field features under the present combustor configuration and injection scheme.

To estimate the order of magnitude of the heat release rate (HRR) in the detonation reaction zone, the thermicity was calculated using the Shock and Detonation Toolbox (SDToolbox; \citealp{kaoSDToolboxNumericalTools}). Representative orders of magnitude for the density, temperature, and specific heat at constant pressure within the reaction zone were then combined with the result to obtain an estimate based on (\ref{eq:10}). For this order-of-magnitude estimate, representative values of the thermicity, density, temperature, and specific heat at constant pressure are taken as \(\dot{\sigma} \sim 10^{6}\,\mathrm{s^{-1}}\), \(\rho \sim 10^{0}\ \mathrm{kg\,m^{-3}}\), \(T \sim 10^{3}\,\mathrm{K}\), and \(c_{p} \sim 10^{3}\ \mathrm{J\,kg^{-1}\,K^{-1}}\), respectively. The resulting order of magnitude of the HRR is therefore \(\text{HRR} \sim 10^{12}\ \mathrm{W\,m^{-3}}\). Accordingly, \(\text{HRR} \geq 10^{12}\,\mathrm{W\,m^{-3}}\) is used as a visualisation threshold to extract regions of intense chemical heat release.

\begin{equation}
\text{HRR} \approx \rho c_{p}T\dot{\sigma}.
\label{eq:10}\end{equation}

This estimate assumes that the thermicity is dominated by the temperature-dependent term and that composition-gradient contributions remain small through the induction and reaction zones; it is used only to set the order-of-magnitude visualisation threshold.

\begin{figure}
\centering
\includegraphics[width=\textwidth,height=0.77\textheight,keepaspectratio]{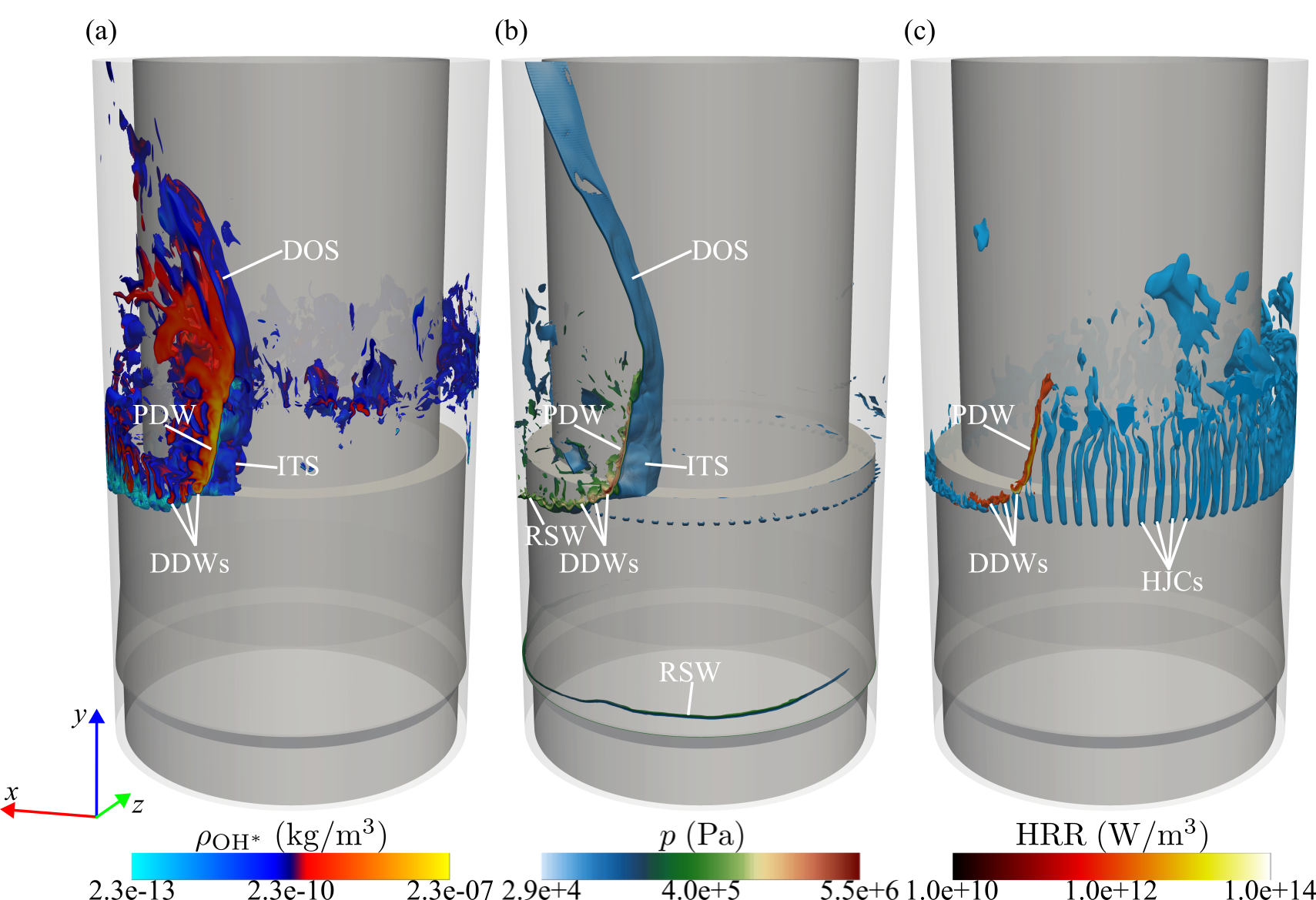}
\caption{Global features of the combustion chamber at \(t = 30\,\mu\mathrm{s}\): (a) OH* density distribution, (b) pressure distribution, and (c) HRR together with hydrogen distribution.}\label{fig:4}
\end{figure}

Figure~\ref{fig:4} presents, in three panels, the characteristic global flow-field structures of the single-wave mode at \(t = 30\,\mu\mathrm{s}\), when the detonation wave is located near \(\theta\) = π. Figure~\ref{fig:4}(a) shows the three-dimensional OH* distribution in the combustor, coloured by OH* density and restricted to the region where \(\left| \nabla\rho_{\text{O}\text{H}^{*}} \right| \geq 7 \times 10^{- 7}\ \mathrm{kg\,m^{-4}}\). Figure~\ref{fig:4}(b) shows the shock-wave system, coloured by static pressure and restricted to the region where \(\lvert\nabla p\rvert/p \geq 1500\,\mathrm{m^{-1}}\). Figure~\ref{fig:4}(c) shows the relationship between the heat-release and species distributions by displaying the region where \(\text{HRR} \geq 10^{12}\ \mathrm{W\,m^{-3}}\) together with the region where \(Y_{H_{2}} \geq 0.1\), shown in light blue.

The principal structures are described below.

The PDW propagates in the circumferential direction. As shown in figure~\ref{fig:4}(b) and figure~\ref{fig:4}(c), the high-heat-release region behind the PDW is concentrated mainly near the outer wall of the combustion section. In the axial direction, the PDW extends approximately from \(y = 64\) to \(84\,\mathrm{mm}\) and does not reach the vicinity of the injection orifices at \(y = 56.78\,\mathrm{mm}\). Thus, it is axially separated from the injector centres and primarily affects the downstream portions of the HJCs.

The DDWs are also concentrated mainly near the outer wall. They extend approximately from \(y = 56.78\) to \(64\,\mathrm{mm}\), covering the upstream portions of the HJCs not directly traversed by the PDW. The similar behaviour of these waves indicates that they continuously separate from the upstream side of the PDW and propagate upstream along the HJCs. Their propagation is therefore primarily axial. During one propagation cycle of the PDW, one DDW can form on each HJC. After propagating past the injection orifice towards the throat, the leading shock becomes decoupled from the heat-release zone, and the DDW degenerates into a reverse shock wave (RSW).

The ITS is distributed mainly near the inner wall of the combustion section and the BFS. Its axial extent spans approximately from the BFS at \(y = 61.8\,\mathrm{mm}\) to \(y = 84\,\mathrm{mm}\). In the direction of detonation-wave propagation, the ITS is slightly phase-advanced relative to the PDW. Its identification in the OH* chemiluminescence images can be attributed to two factors. First, the wave front and the region behind it have relatively high OH* density and therefore emit strongly. Second, because the ITS is phase-advanced relative to the PDW, radial line-of-sight integration produces a discernible luminous structure ahead of the PDW.

The HJCs are fuel-distribution structures formed by the orthogonal interaction between the axial oxidiser crossflow and the radially injected hydrogen jets. Each hydrogen injection orifice produces one corresponding jet column.

\begin{figure}
\centering
\includegraphics[width=\textwidth,height=0.77\textheight,keepaspectratio]{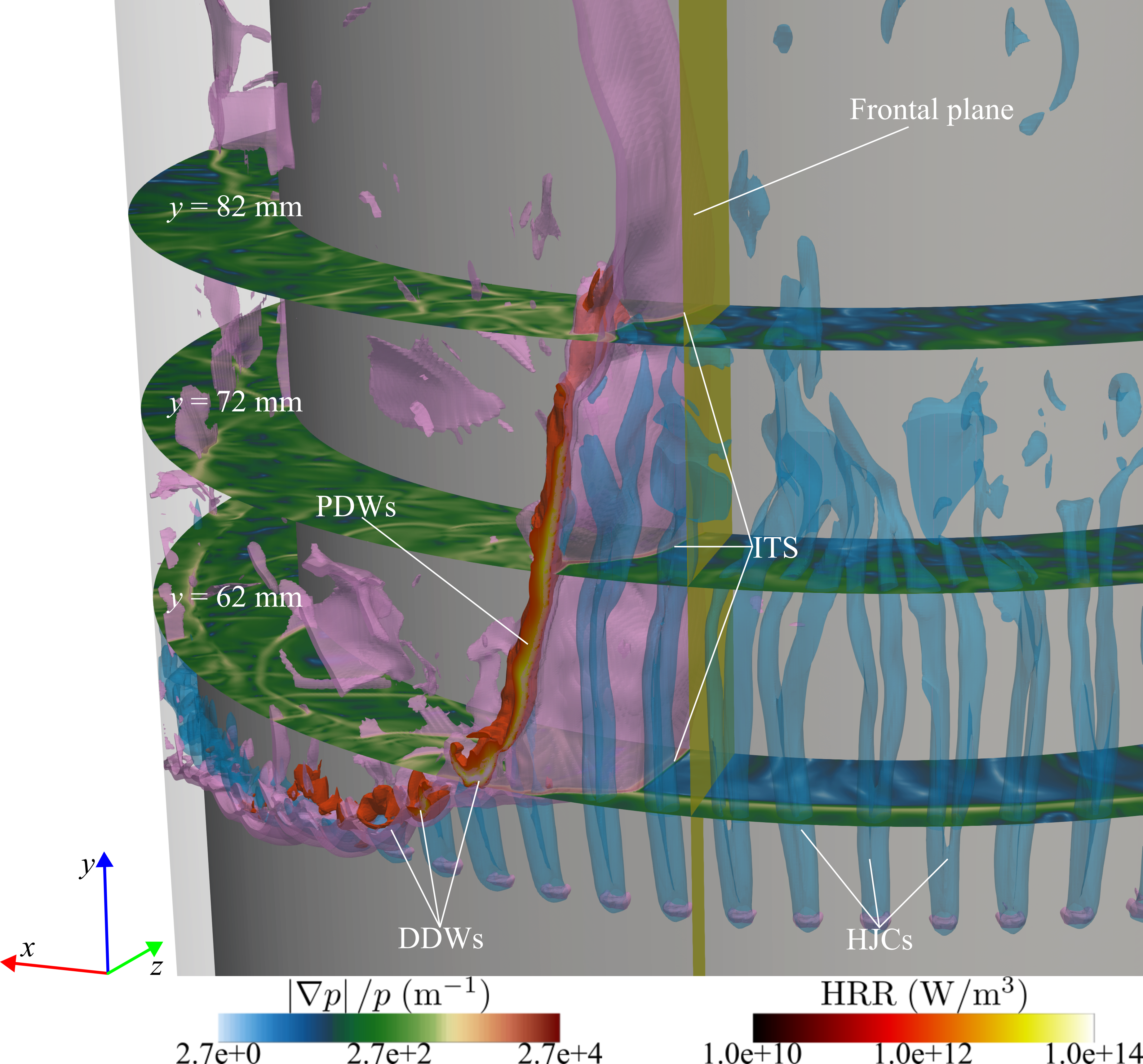}
\caption{Three-dimensional local flow structures associated with the PDW. \(\lvert\nabla p\rvert/p\) contours are shown on three cross-sectional planes at different axial locations. The light-pink region represents the wave structures shown in figure~\ref{fig:4}(b), whereas the light-blue and red regions represent the hydrogen and heat-release distributions shown in figure~\ref{fig:4}(c), respectively. The light-yellow plane denotes the reference plane ahead of the PDW.}\label{fig:5}
\end{figure}

Figure~\ref{fig:5} shows the three-dimensional arrangement of the wave system, heat-release regions and hydrogen distribution near the PDW, highlighting the spatial relationships among the PDW, DDWs, HJCs and ITS. Three cross-sectional planes at different axial locations were selected, and contours of \(\lvert\nabla p\rvert/p\) were plotted on these sections to identify the spatial relationships between the PDW and its associated wave system. The light-pink regions represent the main wave structures shown in figure~\ref{fig:4}(b), while the light-blue and red regions correspond to the hydrogen and heat-release distributions shown in figure~\ref{fig:4}(c), respectively. The light-yellow longitudinal section is located ahead of the PDW and serves as a reference plane for the analysis.

The following sections provide a more detailed analysis and discussion of the PDW, DDWs, and ITS.

\section{Formation mechanisms and interactions of the wave structures}\label{sec:4}

\subsection{Primary detonation wave}\label{sec:4.1}

\begin{figure}
\centering
\includegraphics[width=\textwidth,height=0.77\textheight,keepaspectratio]{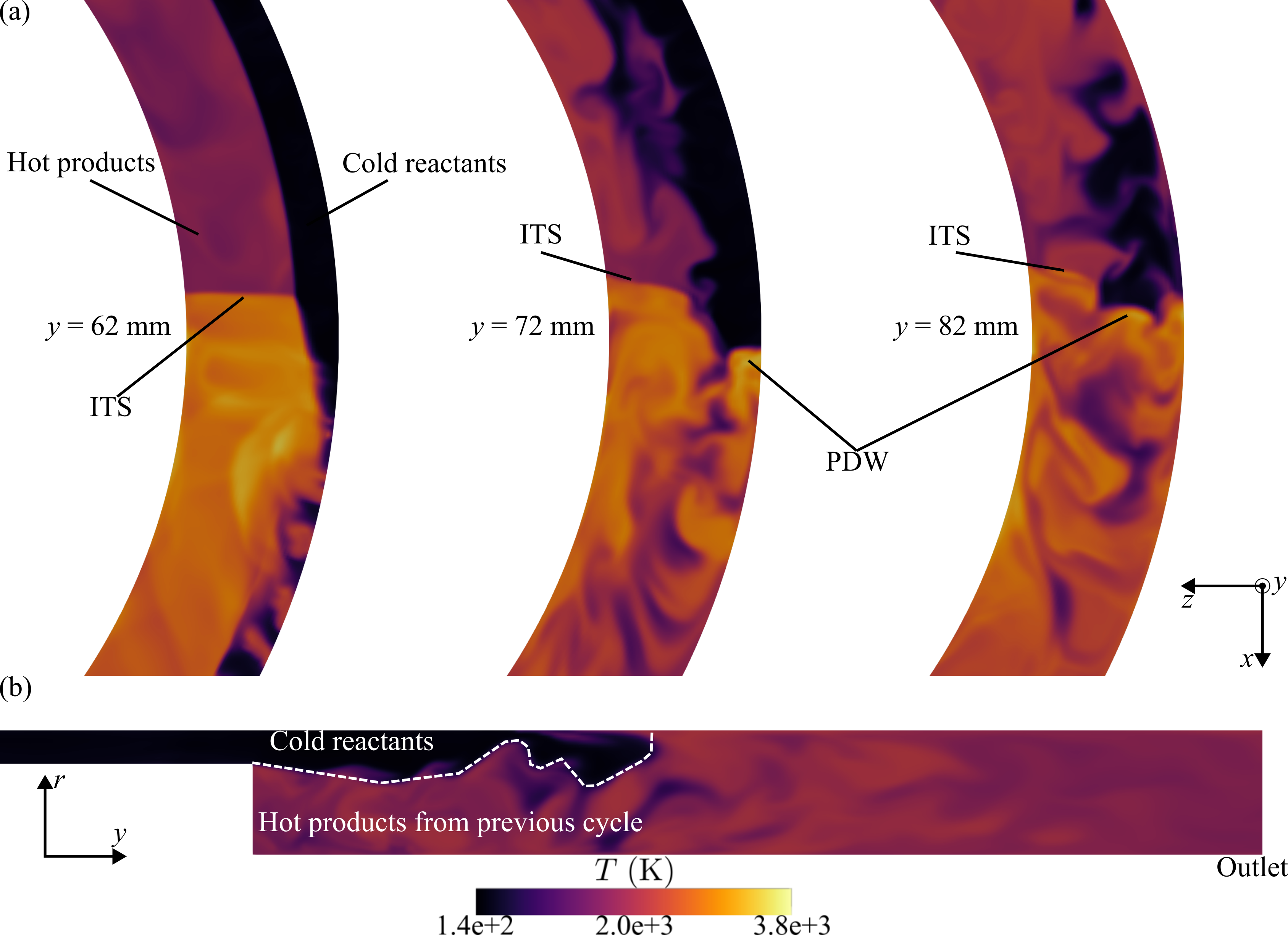}
\caption{Temperature distributions on the planes extracted from figure~\ref{fig:5}: (a) three cross-sectional planes at different axial locations; and (b) the light-yellow reference plane shown in figure~\ref{fig:5}.}\label{fig:6}
\end{figure}

This section focuses on the PDW. Figure~\ref{fig:5} shows that the PDW and its high-heat-release region are concentrated mainly near the outer wall of the combustion section rather than being distributed uniformly throughout the annular passage. To explain this feature, figure~\ref{fig:6}(a) presents the temperature distributions on the three cross-sectional planes shown in figure~\ref{fig:5}. Comparison with the locations of the HJCs and heat-release regions shows that the low-temperature region near the outer wall corresponds to the fresh, unburned mixture ahead of the PDW, whereas the region near the inner wall consists mainly of high-temperature combustion products. Pronounced radial stratification is therefore present. The radial extent of the fresh-mixture layer also closely matches that of the high-heat-release region behind the PDW. Accordingly, the PDW propagates mainly through the fresh-mixture region near the outer wall, and its reaction location is strongly constrained by the local reactant distribution.

Comparison of the three cross-sectional planes shows that, as the flow develops downstream, the interface between the burned products and unburned mixture gradually curves and deforms under the action of shear, revealing pronounced three-dimensional non-uniformity.

To clarify the origin of this radial stratification, figure~\ref{fig:6}(b) presents the temperature distribution on the light-yellow section shown in figure~\ref{fig:5}. The results show that, because of the recirculation region behind the BFS, the high-temperature combustion products generated during the preceding cycle remain concentrated near the inner wall. After injection recovers, the fresh mixture supplied from the mixing section is transported downstream mainly along the outer-wall side, ultimately producing the radially stratified structure.

The PDW propagates predominantly through the fresh-mixture layer near the outer wall, supporting the use of the mean radius of the mixing section, \(r = 36.5\,\mathrm{mm}\), as the reference radius for the circumferential speeds reported in §\,\ref{sec:3.1}.

The Chapman--Jouguet (CJ) velocities were calculated using SDToolbox \citep{kaoSDToolboxNumericalTools} for hydrogen mixed with an oxidiser containing \(40\,\%\) oxygen by volume, at an initial temperature of \(300\,\mathrm{K}\) and an initial pressure of \(1\,\mathrm{atm}\). The CJ velocities were 2066.98 and \(2074.75\,\mathrm{m\,s^{-1}}\) at equivalence ratios of 0.66 and 0.67, respectively. Relative to the corresponding CJ values, the numerical and experimental wave speeds exhibit deficits of approximately \(1.08\,\%\) and \(9.81\,\%\). The numerical result is therefore close to the ideal CJ prediction, whereas the experiment shows a larger velocity deficit. Sub-CJ propagation has also been documented in experimental RDC studies, as summarised in reviews of RDC operation \citep{anandRotatingDetonationCombustors2019,ramanNonidealitiesRotatingDetonation2023}.

\begin{figure}
\centering
\includegraphics[width=\textwidth,height=0.77\textheight,keepaspectratio]{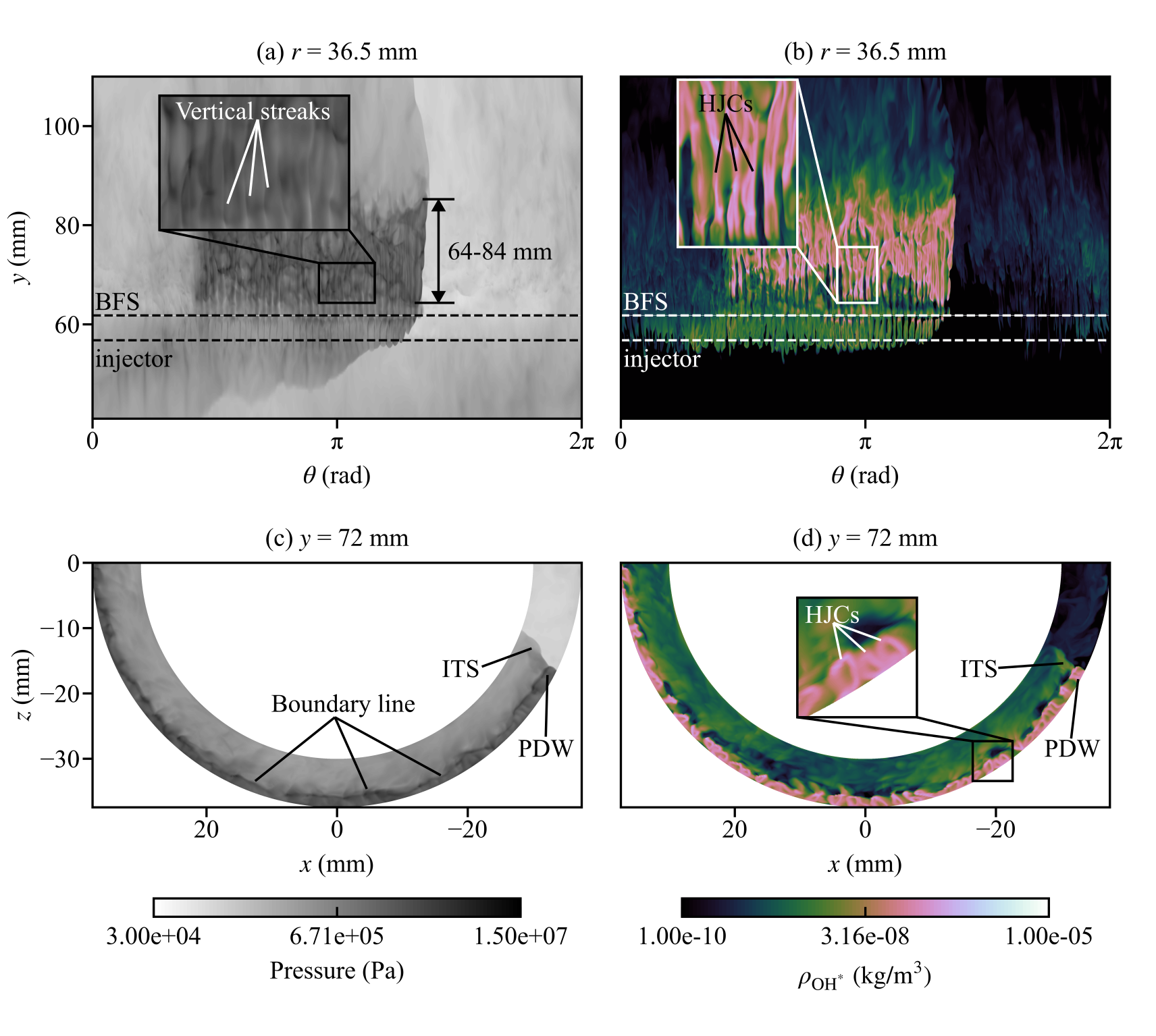}
\caption{Numerical soot-foil maps of the maximum pressure (left column) and maximum OH* density (right column) recorded over \(0\text{–}50\,\mu\mathrm{s}\): (a, b) the cylindrical surface at \(r = 36.5\,\mathrm{mm}\) and (c, d) the cross-sectional plane at \(y = 72\,\mathrm{mm}\).}\label{fig:7}
\end{figure}

To further analyse the propagation characteristics of the PDW, figure~\ref{fig:7} presents numerical soot foils. The plots show the maximum pressure (left column) and maximum OH* density (right column) over \(0\text{–}50\,\mu\mathrm{s}\) on the cylindrical surface at \(r = 36.5\,\mathrm{mm}\) (first row) and the cross-sectional plane (\(y = 72\,\mathrm{mm}\), second row).

As shown in figure~\ref{fig:7}(a), the PDW propagates azimuthally, with its high-pressure region extending approximately from \(y = 64\) to \(84\,\mathrm{mm}\). The close-up view reveals multiple axially elongated streaks, indicating periodic amplification and attenuation of the post-wave pressure as the PDW sweeps across the discrete HJCs. This demonstrates that the non-uniform hydrogen distribution significantly affects the propagation of the PDW. Notably, no distinct diamond-shaped detonation-cell pattern is observed on the selected cylindrical surface over the \(0\text{–}50\,\mu\mathrm{s}\) sampling window.

The regions of high OH* density in figure~\ref{fig:7}(b) correspond closely to the high-pressure regions in figure~\ref{fig:7}(a), further indicating that the PDW and its main reaction zone are concentrated within an axial range extending approximately from \(y = 64\) to \(84\,\mathrm{mm}\). The close-up view shows that the OH* density distribution closely follows the structure of the HJCs, demonstrating that the spatial distribution of the reaction zone is strongly affected by the non-uniform hydrogen distribution.

As shown in figure~\ref{fig:7}(c), a relatively distinct boundary forms between the ITS and the PDW. The high-pressure region is concentrated mainly near the outer wall, further supporting the conclusion that the PDW propagates primarily in the vicinity of the outer wall. The boundary between the two structures is not smooth but exhibits pronounced local fluctuations and deformation. This irregular interface reflects the influence of radial stratification between the burned products and fresh mixture, as well as the morphology of their interface, on the propagation of the PDW and the ITS.

Figure~\ref{fig:7}(d) shows that the high-OH* region is likewise concentrated mainly near the outer wall. The OH* density is higher around the periphery of the HJCs but relatively low inside the columns, producing a distinct annular structure. This occurs because mixing between the outer edges of the HJCs and the surrounding oxidiser is more complete, thereby enhancing the local reaction and heat release.

\subsection{Detached detonation wave}\label{sec:4.2}

This section focuses on the flow structure of the DDWs. As shown in the global flow field in figure~\ref{fig:4} and the local structures in figure~\ref{fig:5}, multiple DDWs exist behind the PDW. Their similar behaviour indicates that they continuously separate from the upstream side of the PDW and propagate upstream along the HJCs. As shown in figure~\ref{fig:5}, the DDWs and their corresponding heat-release zones are also concentrated mainly near the outer wall.

\begin{figure}
\centering
\includegraphics[width=\textwidth,height=0.77\textheight,keepaspectratio]{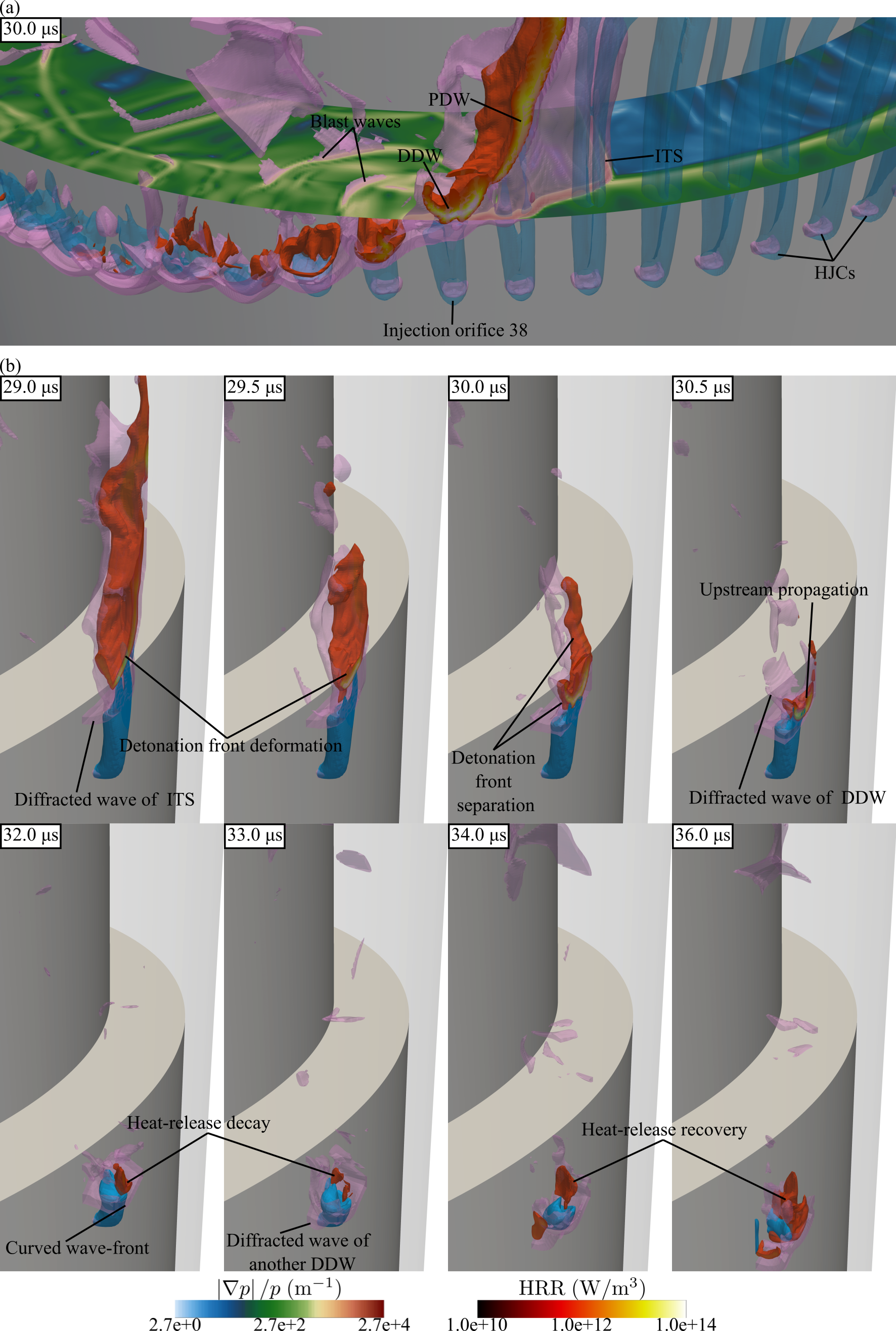}
\caption{Formation and propagation of DDWs: (a) enlarged view of a DDW separating from the PDW; (b) temporal evolution of the DDW in the circumferential sector corresponding to injector orifice No.~38.}\label{fig:8}
\end{figure}

To examine DDW formation, figure~\ref{fig:8}(a) presents a close-up view of the region shown in figure~\ref{fig:5}. A DDW is seen at the stage immediately before separating from the PDW. Based on the similar behaviour of multiple DDWs behind the PDW, their heat-release rates are found to be of the same order of magnitude as that of the PDW. Their leading shocks and post-shock reaction fronts are not only nearly coincident in space but also remain closely coupled from their formation until shock--reaction decoupling begins near the injection orifices. This spatiotemporal consistency indicates clear quasi-detonative characteristics. Given their close connection with the PDW, these structures are defined here as detached detonation waves (DDWs).

To further investigate the formation and propagation of a DDW, figure~\ref{fig:8}(b) presents a time sequence for the circumferential sector (\(\theta \in \left\lbrack \frac{38}{40}\pi,\frac{39}{40}\pi \right\rbrack\)) corresponding to injection orifice No.~38 in figure~\ref{fig:8}(a). The DDW evolution illustrated in figure~\ref{fig:8}(b) is organised into five stages: detonation-front deformation, detonation-front separation, upstream propagation, heat-release decay, and heat-release recovery.

The first stage is detonation-front deformation, corresponding to \(t = 29.0\) and \(29.5\,\mu\mathrm{s}\). Under the influence of the inclination of the PDW, the downstream side of the PDW first separates from the current HJC, while the upstream side remains within it, causing the wave front to curve. As the upstream side gradually separates from the current jet column, the curvature of the wave front increases further.

The second stage is detonation-front separation, corresponding to \(t = 30.0\,\mu\mathrm{s}\). Once the upstream side has completely left the current jet column and entered the neighbouring column, the reaction zone associated with the jet column separates from the PDW, forming a DDW.

The third stage is upstream propagation, illustrated at \(t = 30.5\,\mu\mathrm{s}\). The DDW propagates upstream along the HJC, while its diffracted wave propagates towards the inner wall and eventually forms a blast wave on the BFS, as shown in figure~\ref{fig:8}(a).

The fourth stage is heat-release decay, shown at \(t = 32.0\) and \(33.0\,\mu\mathrm{s}\). As the DDW propagates upstream, the mixing quality becomes poorer and the heat-release zone consequently weakens.

The fifth stage is heat-release recovery, shown at \(t = 34.0\) and \(36.0\,\mu\mathrm{s}\). When the DDW reaches the vicinity of the hydrogen injection orifice, the reaction front expands and combustion surrounds the entire HJC, allowing the heat-release zone to recover gradually.

The formation of DDWs is closely related to the non-uniform mixing of the fresh gases in the combustor. The evolution of the post-wave reaction zone behind the PDW is governed by two competing propagation tendencies. One is circumferential propagation induced by the PDW; the other is upstream propagation along the HJCs, controlled by the local fuel-mixing distribution.

In the upstream portions of the HJCs, poor fuel mixing and the relatively large spacing between adjacent columns make it difficult for the reaction to propagate continuously from one column to the next. Consequently, upstream propagation along the jet columns is dominant. Farther downstream, the fuel becomes better mixed and the spacing between the jet columns decreases, allowing circumferential propagation between adjacent columns to become increasingly dominant. As the dominant propagation tendency changes along the axial direction, the originally connected post-wave reaction front separates: the downstream portion remains coupled to the PDW and continues to propagate circumferentially, while the upstream portion propagates upstream along the HJC and forms a DDW.

\begin{figure}
\centering
\includegraphics[width=\textwidth,height=0.77\textheight,keepaspectratio]{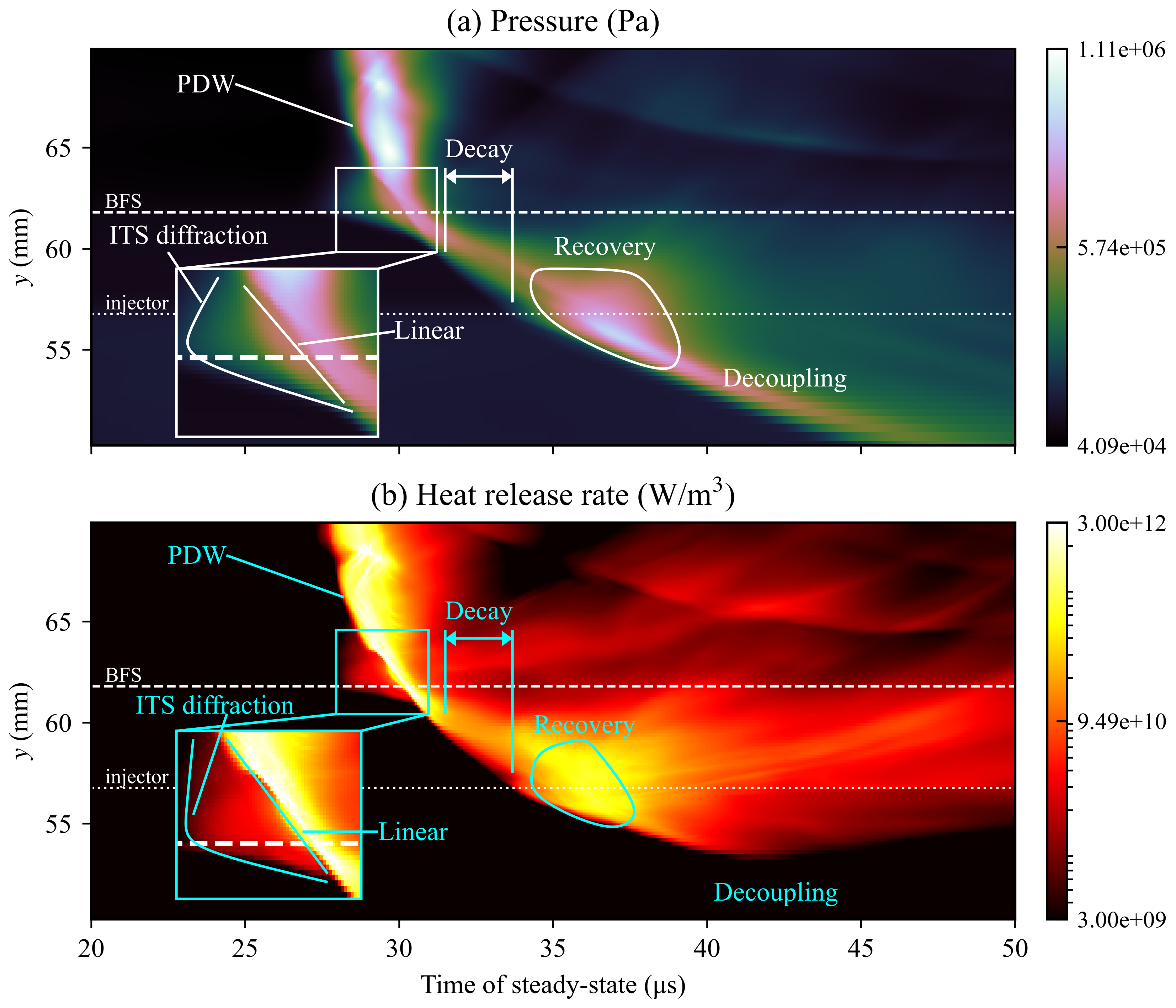}
\caption{Propagation characteristics of the DDW based on area-averaged \(y\)--\(t\) distributions of (a) pressure and (b) HRR in the region corresponding to injector orifice No.~38.}\label{fig:9}
\end{figure}

The DDW propagation characteristics were quantified using pressure and HRR space--time diagrams. The region (\(\theta \in \left\lbrack \frac{38}{40}\pi,\frac{39}{40}\pi \right\rbrack\), \(r \in \lbrack 35.5,37.5\rbrack\text{ mm}\)) associated with injection orifice No.~38 in figure~\ref{fig:8}(a) was area-averaged over the \(\theta\)--\(r\) plane at each axial location, and the resulting \(y\)--\(t\) distributions of pressure and HRR are presented in figure~\ref{fig:9}(a) and figure~\ref{fig:9}(b), respectively.

The pressure and HRR space--time distributions in figure~\ref{fig:9} show that the linear segment used for the speed estimate (\(t \approx 29\text{–}30.5\,\mu\mathrm{s}\)) already spans detonation-front deformation, detonation-front separation, and the early stage of upstream propagation; it is therefore not an additional stage following the first three. Following this linear segment, heat release decays, recovers near the injectors, and eventually decouples from the shock. The close-up view shows a distinct oblique linear pattern in both the pressure and HRR distributions. In this region, the DDW propagates over approximately \(y = 60.9\text{–}63.7\,\mathrm{mm}\) during \(t\) ≈ \(29\text{–}30.5\,\mu\mathrm{s}\). The corresponding absolute propagation speed is approximately \(1866.67\,\mathrm{m\,s^{-1}}\), while its propagation speed relative to the incoming flow is approximately \(2204.67\,\mathrm{m\,s^{-1}}\).

For the corresponding wave-front state, the CJ velocity calculated using \(p = 0.95\,\mathrm{atm}\), \(T = 264\,\mathrm{K}\), and the molar composition \(\mathrm{H_2:O_2:N_2}\) = 0.8:1.0:1.5 is \(1812.53\,\mathrm{m\,s^{-1}}\). Because the propagation speed of the DDW is substantially higher than the corresponding CJ velocity, the wave retains a certain degree of overdriven behaviour for a period immediately after separating from the PDW.

The fourth stage is heat-release decay, corresponding to the weakening of heat release shown in figure~\ref{fig:8}(b). The fifth stage is heat-release recovery, corresponding to the subsequent recovery near the injector in figure~\ref{fig:8}(b). After recovery, the pressure and heat-release signals become decoupled and the DDW degenerates into an RSW.

As established above, the axial extent of the PDW is approximately from \(y = 64\) to \(84\,\mathrm{mm}\), whereas the DDWs propagate upstream over an axial range of approximately \(y = 56.78\) to \(64\,\mathrm{mm}\). Thus, the axial extent of the DDWs corresponds to approximately \(25\text{–}30\,\%\) of the total axial length of the HJCs. This indicates that they are non-negligible heat-release structures in the combustor and have a significant influence on the organisation of the local flow field and the distribution of chemical reactions.

\subsection{Inward-transmitted shock}\label{sec:4.3}

This section focuses on the flow structure of the ITS. As seen in the global flow field in figure~\ref{fig:4} and the local structures in figure~\ref{fig:5}, the ITS is located mainly near the BFS and the inner wall of the combustion section. Compared with the PDW, it is phase-advanced in the direction of propagation and therefore exhibits leading behaviour, giving the overall wave front a three-dimensional geometry with a concave detonation heat-release region.

As shown in figure~\ref{fig:6}, the recirculation region behind the BFS produces pronounced radial temperature stratification in the combustor. The low-temperature region near the outer wall corresponds to the fresh, unburned mixture ahead of the PDW, whereas the region near the inner wall is occupied mainly by high-temperature combustion products. The PDW propagates mainly through the fresh-mixture layer. Its high-pressure disturbance then propagates laterally into the adjacent high-temperature product layer, where it forms a shock structure. From a physical perspective, this structure is a transmitted shock generated when a detonation wave propagates through a radially stratified medium and transmits a disturbance into the lateral inert layer. Because the present combustor employs an inward-expanding configuration, this transmitted shock propagates mainly towards the inner wall and is therefore termed the ``inward-transmitted shock.''

Studies of gaseous layered detonations have shown that the position and propagation state of a transmitted shock relative to the PDW are affected by factors such as the acoustic impedance, sound speed, and area ratio of the two layers \citep{houimInfluenceAcousticImpedance2017,liuRolesAcousticImpedance2026,wahbaDetonationPropagationWeakly2026a}. \citet{wahbaDetonationPropagationWeakly2026a} developed an analytical framework for detonation propagation in a wall-confined reactive--inert two-layer system, relating the propagation regime and coupled wave structure to the acoustic-impedance and layer-area ratios. They formulated a Kantrowitz-type criterion for precursor-shock onset based on choking of the inert layer. The ITS observed here is located ahead of the PDW and exhibits a certain degree of leading propagation. \citet{houimInfluenceAcousticImpedance2017} and \citet{wahbaDetonationPropagationWeakly2026a} referred to transmitted-shock structures with such leading behaviour as a ``detached shock'' and a ``precursor shock,'' respectively. These studies provide useful theoretical references for understanding the leading propagation of transmitted shocks in stratified media.

\begin{figure}
\centering
\includegraphics[width=\textwidth,height=0.77\textheight,keepaspectratio]{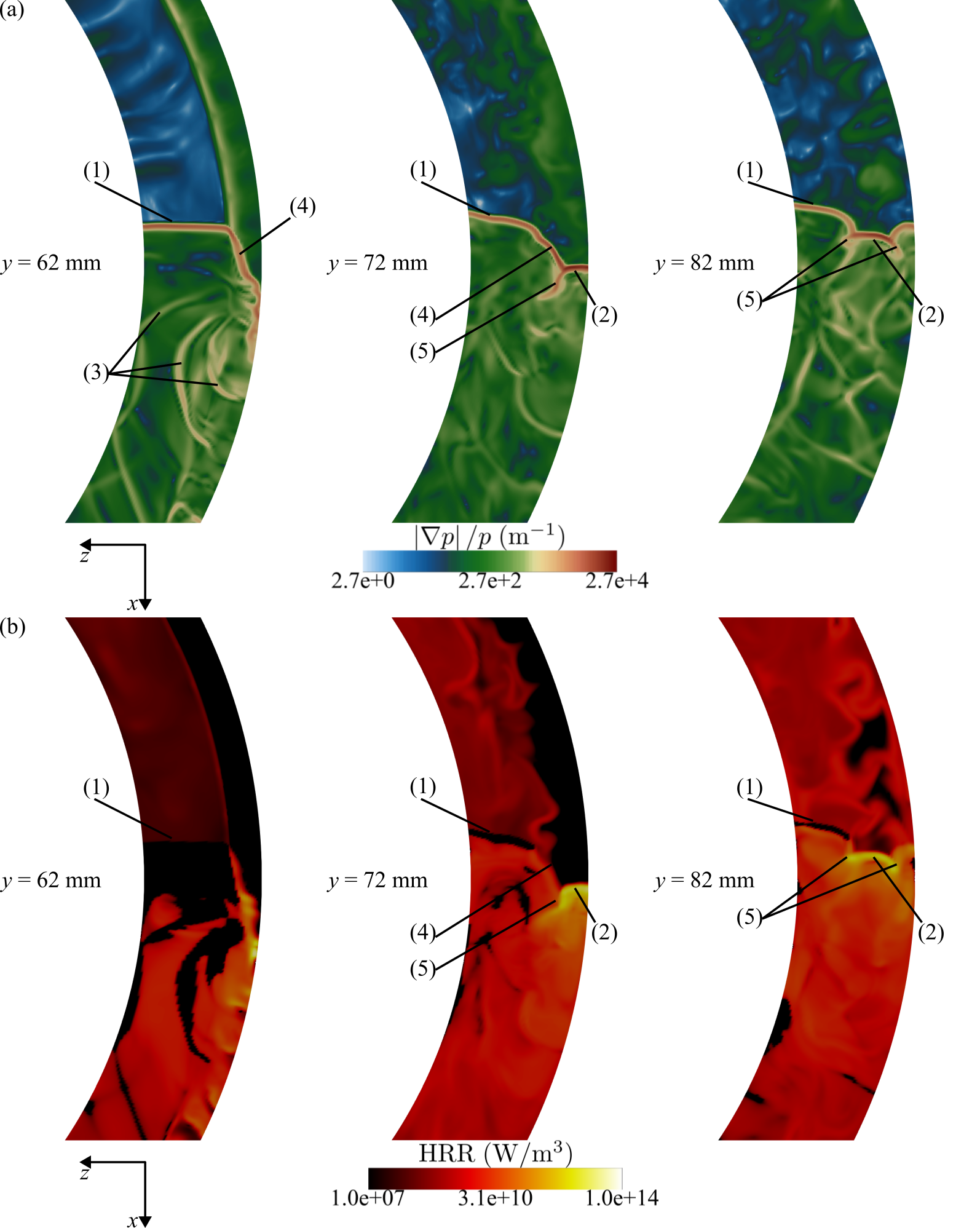}
\caption{Two-dimensional wave structures and heat-release-rate distributions on the three cross-sectional planes extracted from figure~\ref{fig:5}: (a) normalised pressure-gradient magnitude; and (b) heat-release-rate distribution. The labelled wave structures are (1) the ITS, (2) the PDW, (3) the blast waves, (4) the transmitted oblique shock, and (5) the reflected oblique shock.}\label{fig:10}
\end{figure}

To further analyse the two-dimensional spatial relationship between the ITS and the PDW in the radially stratified medium, figure~\ref{fig:10}(a) and figure~\ref{fig:10}(b) show the \(\lvert\nabla p\rvert/p\) and HRR distributions, respectively, on the three cross-sectional planes marked in figure~\ref{fig:5}. In the figure, (1) denotes the ITS, (2) the PDW, (3) the blast waves, (4) the transmitted oblique shock, and (5) the reflected oblique shock. The terms ``transmitted oblique shock'' and ``reflected oblique shock'' follow the description of the wave system associated with the detached-shock configuration in gaseous layered detonations given by \citet{houimInfluenceAcousticImpedance2017}.

Figure~\ref{fig:10}(a) shows that the ITS propagates into the fresh-mixture region near the outer wall and induces a transmitted oblique shock. At \(y = 62\,\mathrm{mm}\), no PDW propagates circumferentially across the section, allowing the transmitted oblique shock to extend to the outer wall. At \(y = 72\,\mathrm{mm}\), the transmitted oblique shock interacts with the PDW and forms a reflected oblique shock. At \(y = 82\,\mathrm{mm}\), the wave system becomes even more complex.

Figure~\ref{fig:10}(b) shows a local reduction in the net HRR near the front of the ITS, whereas enhanced HRR occurs near the transmitted oblique shock, reflected oblique shock, and PDW. This is because the ITS propagates mainly through the high-temperature combustion-product layer. Its high-temperature and high-pressure conditions alter the local reaction balance, causing the reactions to be dominated by heat absorption. This feature clearly distinguishes the ITS from the other wave structures.

Together with figure~\ref{fig:6}, figure~\ref{fig:10} suggests that the leading behaviour of the ITS may be jointly affected by radial medium stratification and the curvature of the annular walls. In the present combustor, the relevant stratification is radial: hot combustion products occupy the region near the inner wall, while fresh reactants are concentrated near the outer wall. The ITS propagates through the hot-product layer ahead of the PDW in circumferential phase. For comparison with the two-layer model of \citet{wahbaDetonationPropagationWeakly2026a}, the hot-product and fresh-reactant layers are associated with the inert and reactive layers, respectively, with the inner and outer walls providing radial confinement. The layer-area ratio is estimated as approximately 2.75 from the radial extents of the two layers at \(y = 72\,\mathrm{mm}\); the acoustic-impedance ratio is estimated as approximately 0.35 using the mean temperatures of the two layers. For this area ratio, the critical acoustic-impedance ratio for inert-layer choking is approximately 0.40 (read from figure~8 of \citealp{wahbaDetonationPropagationWeakly2026a}); the present estimate (approximately 0.35) lies below this estimated critical value. This indicates that, at the theoretical level, the model can partially explain the precursor behaviour.

Wall curvature may also influence this spatial relationship. In studies of gaseous layered detonations, the transmitted shock and the detonation wave propagate along straight paths, with similar linear velocities and a relatively stable separation distance. In the present annular combustor, however, both waves propagate circumferentially. Their angular velocities can therefore remain approximately equal, allowing the phase difference between them to remain relatively stable within a certain range. The annular curvature may thus help maintain the phase-advanced position of the ITS relative to the PDW.

In the \(y = 62\,\mathrm{mm}\) section shown in figure~\ref{fig:10}(a), a series of periodic blast waves can be observed behind the ITS. Because no PDW propagates circumferentially across this section, the waves crossing the section are primarily DDWs. This observation suggests that the PDW and DDWs may participate in the formation and maintenance of the ITS through different pathways. In the PDW region, the ITS may be sustained mainly by continuous high-pressure disturbances, whereas in the DDW region, it may receive energy primarily from discrete blast waves that catch up with it as they propagate inward.

\begin{figure}
\centering
\includegraphics[width=\textwidth,height=0.77\textheight,keepaspectratio]{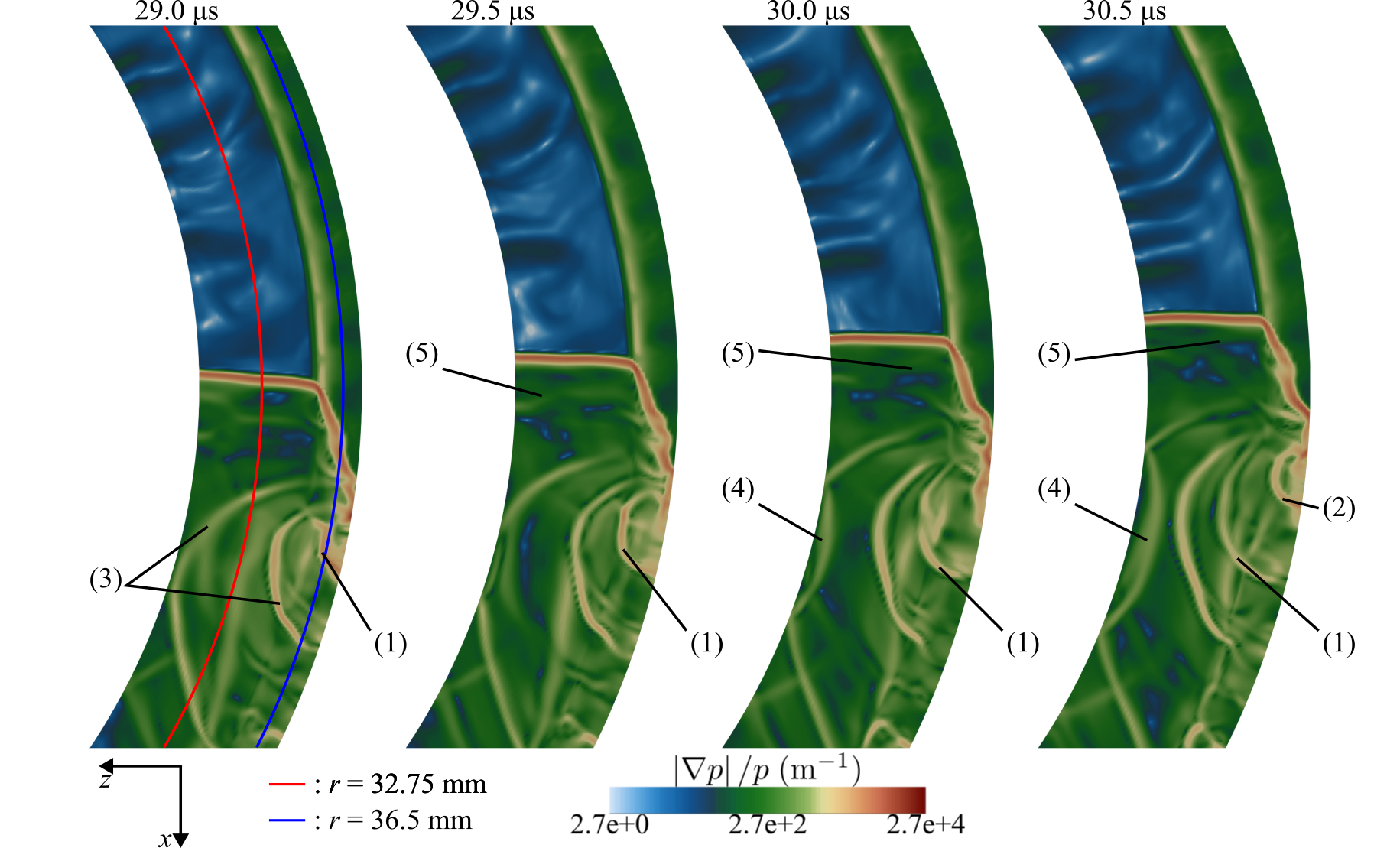}
\caption{Temporal evolution of blast-wave formation and interaction with the ITS on the \(y = 62\,\mathrm{mm}\) cross-section at four successive instants. The labelled structures are (1) formation and propagation of a blast wave, (2) formation of a new blast wave, (3) other blast waves, (4) reflected waves generated by the blast wave, and (5) a blast wave catching up with the ITS.}\label{fig:11}
\end{figure}

To further examine the formation of these blast waves and their interaction with the ITS, figure~\ref{fig:11} presents the instantaneous \(\lvert\nabla p\rvert/p\) distributions at four time instants on the \(y = 62\,\mathrm{mm}\) cross-section. In the figure, (1) denotes the formation and propagation of a blast wave, (2) the formation of a new blast wave, (3) other blast waves, (4) reflected blast waves, and (5) the process by which a blast wave catches up with the ITS.

Observations of the evolution of (1) and (2) show that blast waves are generated periodically at the outer wall and propagate towards the inner wall, at a speed substantially higher than that of the ITS. Further examination of (3) and (4) indicates that the complex wave system behind the ITS consists primarily of multiple blast waves and their reflected waves. As shown in (5), these blast waves gradually catch up with the ITS during propagation and eventually merge with it.

\begin{figure}
\centering
\includegraphics[width=\textwidth,height=0.77\textheight,keepaspectratio]{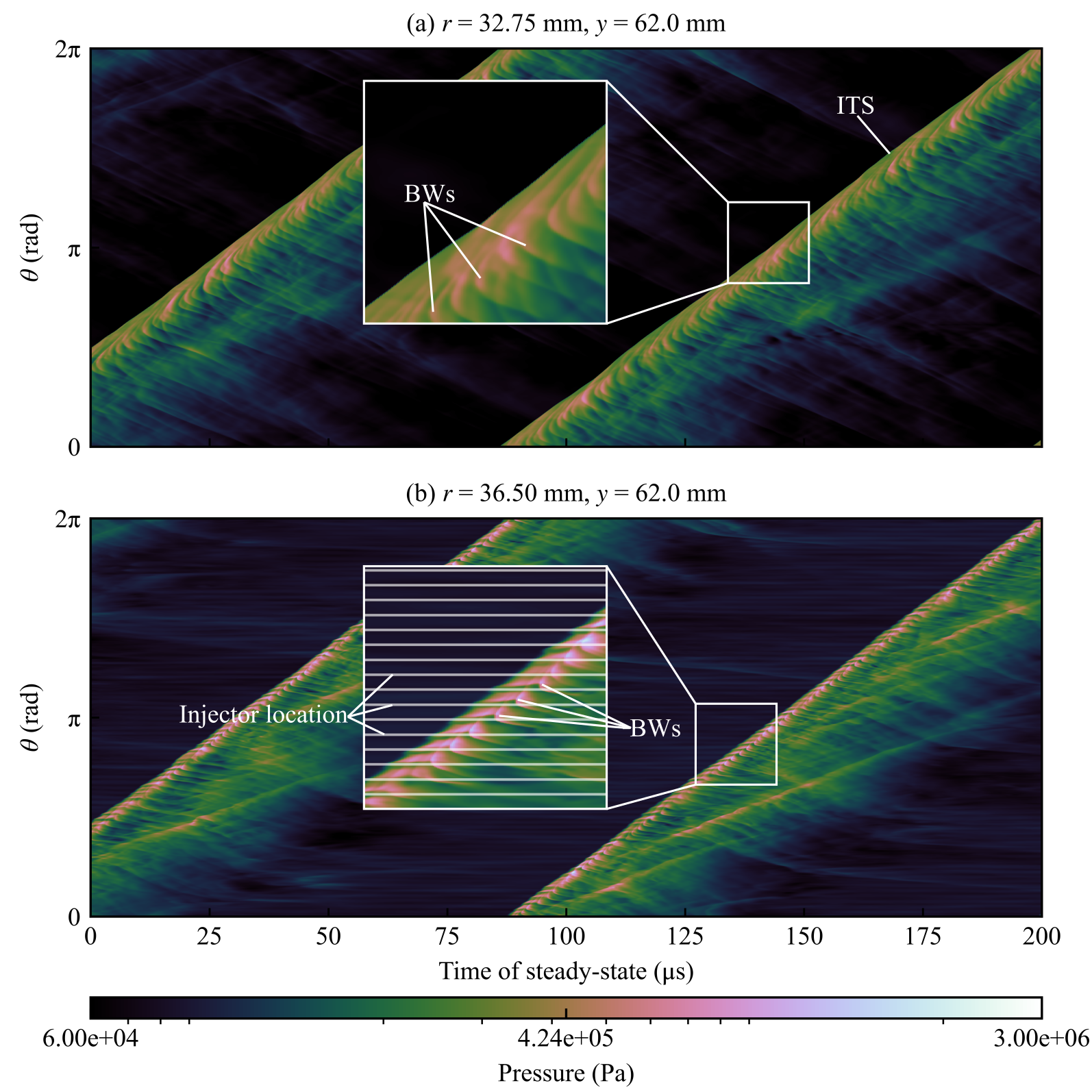}
\caption{Pressure space--time diagrams along two circumferential lines over a \(200\,\mu\mathrm{s}\) time window: (a) \(y = 62\,\mathrm{mm}\), \(r = 32.75\,\mathrm{mm}\), within the ITS region; (b) \(y = 62\,\mathrm{mm}\), at the reference radius \(r = 36.5\,\mathrm{mm}\).}\label{fig:12}
\end{figure}

To track the process by which the blast waves catch up with the ITS and to investigate their possible origin, figure~\ref{fig:12} presents \(\theta\)--\(t\) diagrams of the pressure along two circumferential lines shown in figure~\ref{fig:11} over a \(0\text{–}200\,\mu\mathrm{s}\) time window. Figure~\ref{fig:12}(a) corresponds to the red line crossing the ITS region on the \(y = 62\,\mathrm{mm}\) cross-section, at a radial position of \(r = 32.75\,\mathrm{mm}\). Figure~\ref{fig:12}(b) corresponds to the blue line on the same cross-section at the reference radius of \(r = 36.5\,\mathrm{mm}\).

The enlarged view in figure~\ref{fig:12}(a) shows that blast waves are periodically generated behind the ITS in the detached-detonation-wave region. These blast waves propagate substantially faster than the ITS and gradually catch up with it. Their trajectories ultimately terminate near the shock front, where they merge with the ITS. These periodic blast waves provide recurring pressure disturbances that may help maintain a comparatively steady propagation velocity of the ITS.

The enlarged view in figure~\ref{fig:12}(b) and the snapshots in figure~\ref{fig:11} show that the blast waves propagating inward from the vicinity of the outer wall exhibit a distinct periodicity. Their generation phases correspond one-to-one with the circumferential positions of the hydrogen injection orifices. This indicates that the blast waves on the \(y = 62\,\mathrm{mm}\) cross-section are not generated randomly. Instead, they are formed as the diffracted shocks of DDWs propagate towards the BFS region while the DDWs travel upstream along the HJCs. Their three-dimensional structure can be seen in figure~\ref{fig:8}.

These observations indicate that pressure disturbances generated during DDW evolution contribute to maintaining the ITS in the DDW region. In particular, blast waves generated by diffraction of the DDW leading shocks propagate inward and catch up with the ITS, providing additional pressure disturbances that support its propagation.

\section{Conclusions}\label{sec:5}

Three-dimensional, unsteady large-eddy simulations of a non-premixed hydrogen--air rotating detonation combustor were performed using JANC, a JAX-based compressible reactive-flow solver. The structural characteristics and formation mechanisms of the detonation wave and its associated wave system were systematically investigated. The main conclusions are as follows:

\begin{enumerate}
\def\labelenumi{(\arabic{enumi})}
\tightlist
\item
  Under the representative operating condition, the simulation captured the experimentally observed single-wave mode, nearly periodic pressure fluctuations, double-peaked phase-averaged pressure waveform, and major OH* emission features. The predicted mean pressure-cycle period was approximately \(8.48\,\%\) shorter than the experimental value. Comparison of the experimental and simulated results indicates that the model captures the main pressure-fluctuation and emission characteristics under this condition and can be used to analyse the three-dimensional wave system in the combustor.
\end{enumerate}

\begin{enumerate}
\def\labelenumi{(\arabic{enumi})}
\setcounter{enumi}{1}
\tightlist
\item
  Recirculation induced by the BFS concentrated high-temperature combustion products near the inner wall, while fresh reactants were distributed mainly near the outer wall. Consequently, the PDW and its main heat-release region propagated predominantly near the outer wall. The non-uniform fuel distribution associated with the discrete HJCs affected the propagation extent of the PDW and produced periodic variations in local pressure and reaction intensity.
\end{enumerate}

\begin{enumerate}
\def\labelenumi{(\arabic{enumi})}
\setcounter{enumi}{2}
\tightlist
\item
  Multiple DDWs successively separated from the upstream side of the PDW and propagated upstream along the HJCs. Their leading shocks remained closely coupled to the reaction zones, and their heat-release rates were of the same order of magnitude as that of the PDW. Immediately after separation, the wave fronts briefly propagated faster than the corresponding CJ velocity. The DDWs formed through competition between circumferential propagation driven by the PDW and upstream propagation supported by the local fuel distribution. DDW evolution proceeded through detonation-front deformation, detonation-front separation, upstream propagation along the HJCs, heat-release decay, and heat-release recovery near the injectors, followed by eventual shock--reaction decoupling.
\end{enumerate}

\begin{enumerate}
\def\labelenumi{(\arabic{enumi})}
\setcounter{enumi}{3}
\tightlist
\item
  The ITS was located mainly near the BFS and the inner wall, and was phase-advanced relative to the PDW in the circumferential direction. The results suggest that radial stratification, annular-wall curvature, PDW-induced pressure disturbances and periodic blast waves generated during DDW evolution jointly influence ITS propagation. Although the ITS was not itself a major heat-release structure, it induced transmitted oblique shocks in the fresh-mixture region near the outer wall and formed reflected oblique shocks through interaction with the PDW. In regions dominated by DDWs, blast waves were generated periodically near the HJCs and propagated towards the inner wall. Their generation phase corresponded to the circumferential arrangement of the injection orifices, indicating that they were primarily induced by diffraction of the DDW leading shocks as they propagated inward. The blast waves travelled faster than the ITS and eventually caught up with it and merged with it. The resulting pressure disturbances contributed to maintaining the propagation of the ITS.
\end{enumerate}

Overall, the simulations helped identify the luminous structures observed in the experimental OH* snapshots, particularly the DDWs and ITS, and provided further insight into their structures, propagation characteristics and formation mechanisms within the rotating detonation combustor.
\begin{bmhead}[Funding.]
The authors would like to acknowledge support from the National Natural Science Foundation of China (grant nos. 52306152 and 12441201).
\end{bmhead}

\begin{bmhead}[Declaration of interests.]
The authors report no conflict of interest.
\end{bmhead}

\begin{bmhead}[Data availability statement.]
The JANC solver is openly available at \url{https://github.com/AllinSim/JANC/}. The implementation specific to the present RDC simulations and the flow-field data that support the findings of this study are available from the corresponding author upon reasonable request.
\end{bmhead}

\begin{bmhead}[Use of AI tools.]
During the preparation of this work, the authors used AI-assisted translation and language polishing to improve the English expression of the manuscript. The authors reviewed and edited the content and take full responsibility for the publication.
\end{bmhead}

\begin{bmhead}[Author contributions.]
Ruiming Cao: Conceptualization, Data curation, Formal analysis, Investigation, Methodology, Project administration, Resources, Software, Supervision, Validation, Visualization, Writing--original draft, Writing--review \& editing. Faxuan Luo: Writing--review \& editing. Mingyang Bu: Writing--review \& editing. Yingchen Shi: Writing--review \& editing. Haocheng Wen: Funding acquisition, Resources, Validation, Writing--review \& editing. Bing Wang: Funding acquisition, Resources, Validation, Writing--review \& editing.
\end{bmhead}

\bibliographystyle{jfm}
\bibliography{references}

\begin{thebibliography}{57}
\expandafter\ifx\csname natexlab\endcsname\relax\def\natexlab#1{#1}\fi
\def\au#1{#1} \def\ed#1{#1} \def\yr#1{#1}\def\at#1{#1}\def\jt#1{\textit{#1}}
  \def\bt#1{#1}\def\bvol#1{\textbf{#1}} \def\vol#1{#1} \def\pg#1{#1}
  \def\publ#1{#1}\def\arxiv#1{#1}\def\org#1{#1}\def\st#1{\textit{#1}}

\bibitem[Anand \& Gutmark(2019)]{anandRotatingDetonationCombustors2019}
{\sc \au{Anand, V.} \& \au{Gutmark, E.}} \yr{2019}  \at{Rotating detonation
  combustors and their similarities to rocket instabilities}.  \jt{Prog. Energy
  Combust. Sci.}  \bvol{73},  \pg{182--234}.

\bibitem[Armbruster {\em et~al.\/}(2026)Armbruster, Bard, Börner, Rheindorf,
  Martin, Lober, General \& Hardi]{armbrusterVisualizationDetonationWave2026}
{\sc \au{Armbruster, W.}, \au{Bard, E.}, \au{Börner, M.}, \au{Rheindorf, K.},
  \au{Martin, J.}, \au{Lober, L.}, \au{General, S.} \& \au{Hardi, J.~S.}}
  \yr{2026}  \at{Visualization of the detonation wave structure in a
  small-scale hydrogen–oxygen rotating detonation combustor}.  \jt{Aerospace}
   \bvol{13}~(2),  \pg{158}.

\bibitem[Athmanathan {\em et~al.\/}(2022)Athmanathan, Braun, Ayers, Fugger,
  Webb, Slipchenko, Paniagua, Roy \&
  Meyer]{athmanathanEffectsReactantStratification2022}
{\sc \au{Athmanathan, V.}, \au{Braun, J.}, \au{Ayers, Z.~M.}, \au{Fugger,
  C.~A.}, \au{Webb, A.~M.}, \au{Slipchenko, M.~N.}, \au{Paniagua, G.}, \au{Roy,
  S.} \& \au{Meyer, T.~R.}} \yr{2022}  \at{On the effects of reactant
  stratification and wall curvature in non-premixed rotating detonation
  combustors}.  \jt{Combust. Flame}  \bvol{240},  \pg{112013}.

\bibitem[Bluemner {\em et~al.\/}(2019)Bluemner, Bohon, Paschereit \&
  Gutmark]{bluemnerExperimentalStudyReactant2019}
{\sc \au{Bluemner, R.}, \au{Bohon, M.~D.}, \au{Paschereit, C.~O.} \&
  \au{Gutmark, E.~J.}} \yr{2019}  \at{Experimental study of reactant mixing in
  model rotating detonation combustor geometries}.  \jt{Flow Turbul. Combust.}
  \bvol{102}~(2),  \pg{255--277}.

\bibitem[Bonanni {\em et~al.\/}(2024)Bonanni, Brouzet, Vignat \&
  Ihme]{bonanniCouplingDetonationStructure2024a}
{\sc \au{Bonanni, M.}, \au{Brouzet, D.}, \au{Vignat, G.} \& \au{Ihme, M.}}
  \yr{2024}  \at{Coupling of detonation structure and upstream inhomogeneities
  in a rotating detonation engine}.  \jt{Proc. Combust. Inst.}
  \bvol{40}~(1-4),  \pg{105576}.

\bibitem[Bu {\em et~al.\/}(2026)Bu, Wen, Dai, Xie, Shi \&
  Wang]{buRotatingDetonationBladeless2026}
{\sc \au{Bu, M.}, \au{Wen, H.}, \au{Dai, W.}, \au{Xie, Q.}, \au{Shi, Y.} \&
  \au{Wang, B.}} \yr{2026}  \at{Rotating detonation bladeless turbine
  prototype: An experimental analysis of combustion characteristics and power
  output}.  \jt{Aerosp. Sci. Technol.}  \bvol{178},  \pg{112401}.

\bibitem[Burke {\em et~al.\/}(2021)Burke, Rezzag, Dunn, Flores \&
  Ahmed]{burkeEffectPremixedStratification2021}
{\sc \au{Burke, R.}, \au{Rezzag, T.}, \au{Dunn, I.}, \au{Flores, W.} \&
  \au{Ahmed, K.}} \yr{2021}  \at{The effect of premixed stratification on the
  wave dynamics of a rotating detonation combustor}.  \jt{Int. J. Hydrogen
  Energy}  \bvol{46}~(54),  \pg{27816--27826}.

\bibitem[Chacon {\em et~al.\/}(2021)Chacon, Feleo \&
  Gamba]{chaconSecondaryWavesDynamics2021}
{\sc \au{Chacon, F.}, \au{Feleo, A.~D.} \& \au{Gamba, M.}} \yr{2021}
  \at{Secondary waves dynamics and their impact on detonation structure in
  rotating detonation combustors}.  \jt{Shock Waves}  \bvol{31}~(7),
  \pg{675--702}.

\bibitem[Chen {\em et~al.\/}(2026)Chen, Liu, Peng, Fan, Yang, Yan, Dai \&
  Liu]{chenEffectsBackwardfacingStep2026}
{\sc \au{Chen, Y.}, \au{Liu, S.}, \au{Peng, H.}, \au{Fan, W.}, \au{Yang, H.},
  \au{Yan, C.}, \au{Dai, Y.} \& \au{Liu, W.}} \yr{2026}  \at{Effects of
  backward-facing step depths and inner/outer-wall injection schemes on
  rotating detonation wave under high flight {Mach} number conditions}.
  \jt{Energy}  \bvol{344},  \pg{140012}.

\bibitem[Cho {\em et~al.\/}(2016)Cho, Codoni, Rankin, Hoke \&
  Schauer]{choHighRepetitionRateChemiluminescenceImaging2016}
{\sc \au{Cho, K.~Y.}, \au{Codoni, J.~R.}, \au{Rankin, B.~A.}, \au{Hoke, J.~L.}
  \& \au{Schauer, F.~R.}} \yr{2016} High-repetition-rate chemiluminescence
  imaging of a rotating detonation engine.  \bt{In {\em 54th {AIAA} {Aerospace}
  {Sciences} {Meeting}\/}}.  \publ{San Diego, California, USA: American
  Institute of Aeronautics and Astronautics}, {AIAA} Paper 2016-1648.

\bibitem[Cocks {\em et~al.\/}(2016)Cocks, Holley \&
  Rankin]{cocksHighFidelitySimulations2016}
{\sc \au{Cocks, P. A.~T.}, \au{Holley, A.~T.} \& \au{Rankin, B.~A.}} \yr{2016}
  High fidelity simulations of a non-premixed rotating detonation engine.
  \bt{In {\em 54th {AIAA} {Aerospace} {Sciences} {Meeting}\/}}.  \publ{San
  Diego, California, USA: American Institute of Aeronautics and Astronautics},
  {AIAA} Paper 2016-0125.

\bibitem[Dai {\em et~al.\/}(2026)Dai, Shi, Ma, Bu, Wang, Xie, Wen \&
  Wang]{daiHighPerformanceUltravioletOptical2026}
{\sc \au{Dai, W.}, \au{Shi, Y.}, \au{Ma, J.}, \au{Bu, M.}, \au{Wang, L.},
  \au{Xie, Q.}, \au{Wen, H.} \& \au{Wang, B.}} \yr{2026}  \at{A
  high-performance ultraviolet optical sensing system for rotating detonation
  extreme combustion}.  \jt{Sensors}  \bvol{26}~(10),  \pg{3248}.

\bibitem[{EDL}(2023)]{kaoSDToolboxNumericalTools}
{\sc \au{{EDL}}} \yr{2023}  \bt{{SDToolbox}: Numerical tools for shock and
  detonation wave modeling}. {GALCIT} {Report} FM2018.001.  \org{Explosion
  Dynamics Laboratory, California Institute of Technology}, Pasadena, CA.

\bibitem[Fievisohn \& Yu(2017)]{fievisohnSteadyStateAnalysisRotating2017}
{\sc \au{Fievisohn, R.~T.} \& \au{Yu, K.~H.}} \yr{2017}  \at{Steady-state
  analysis of rotating detonation engine flowfields with the method of
  characteristics}.  \jt{J. Propul. Power}  \bvol{33}~(1),  \pg{89--99}.

\bibitem[Fugger {\em et~al.\/}(2021)Fugger, Lopez, Rein, Roy \&
  Caswell]{fuggerDynamicsNonpremixedRotating2021}
{\sc \au{Fugger, C.~A.}, \au{Lopez, J.~G.}, \au{Rein, K.~D.}, \au{Roy, S.} \&
  \au{Caswell, A.~W.}} \yr{2021}  \at{The dynamics of a non-premixed rotating
  detonation engine from time-resolved temperature measurements}.  \jt{Proc.
  Combust. Inst.}  \bvol{38}~(3),  \pg{3787--3795}.

\bibitem[Fujii {\em et~al.\/}(2017)Fujii, Kumazawa, Matsuo, Nakagami, Matsuoka
  \& Kasahara]{fujiiNumericalInvestigationDetonation2017}
{\sc \au{Fujii, J.}, \au{Kumazawa, Y.}, \au{Matsuo, A.}, \au{Nakagami, S.},
  \au{Matsuoka, K.} \& \au{Kasahara, J.}} \yr{2017}  \at{Numerical
  investigation on detonation velocity in rotating detonation engine chamber}.
  \jt{Proc. Combust. Inst.}  \bvol{36}~(2),  \pg{2665--2672}.

\bibitem[Giassi {\em et~al.\/}(2016)Giassi, Cao, Bennett, Stocker, Takahashi,
  Smooke \& Long]{giassiAnalysisCHConcentration2016}
{\sc \au{Giassi, D.}, \au{Cao, S.}, \au{Bennett, B. A.~V.}, \au{Stocker,
  D.~P.}, \au{Takahashi, F.}, \au{Smooke, M.~D.} \& \au{Long, M.~B.}} \yr{2016}
   \at{Analysis of {CH}* concentration and flame heat release rate in laminar
  coflow diffusion flames under microgravity and normal gravity}.  \jt{Combust.
  Flame}  \bvol{167},  \pg{198--206}.

\bibitem[Hellard {\em et~al.\/}(2024)Hellard, Gaillard, Davidenko,
  Berterretche, Zitoun \& Vidal]{hellardQuasiCJRotatingDetonation2024}
{\sc \au{Hellard, P.}, \au{Gaillard, T.}, \au{Davidenko, D.}, \au{Berterretche,
  P.}, \au{Zitoun, R.} \& \au{Vidal, P.}} \yr{2024}  \at{Quasi-{CJ} rotating
  detonation with partially premixed methane-oxygen injection: Numerical
  simulation and experimental validation}.  \jt{Appl. Energy Combust. Sci.}
  \bvol{19},  \pg{100278}.

\bibitem[Hishida {\em et~al.\/}(2009)Hishida, Fujiwara \&
  Wolanski]{hishidaFundamentalsRotatingDetonations2009}
{\sc \au{Hishida, M.}, \au{Fujiwara, T.} \& \au{Wolanski, P.}} \yr{2009}
  \at{Fundamentals of rotating detonations}.  \jt{Shock Waves}  \bvol{19}~(1),
  \pg{1--10}.

\bibitem[Hishida \& Hayashi(1993)]{hishidaNumericalSimulationPulsed1993}
{\sc \au{Hishida, M.} \& \au{Hayashi, A.~K.}} \yr{1993} Numerical simulation of
  pulsed jet plume combustion.  \bt{In {\em Dynamics of Heterogeneous
  Combustion and Reacting Systems\/}},  \st{Progress in Astronautics and
  Aeronautics},  \vol{vol. 152},  \pg{pp. 343--361}.  \publ{Washington, DC:
  American Institute of Aeronautics and Astronautics}.

\bibitem[Houim \& Fievisohn(2017)]{houimInfluenceAcousticImpedance2017}
{\sc \au{Houim, R.~W.} \& \au{Fievisohn, R.~T.}} \yr{2017}  \at{The influence
  of acoustic impedance on gaseous layered detonations bounded by an inert
  gas}.  \jt{Combust. Flame}  \bvol{179},  \pg{185--198}.

\bibitem[Huang {\em et~al.\/}(2023{\natexlab{{\em a\/}}})Huang, Zhou, Liu,
  Peng, Yuan \& Zhang]{huangPropagationFlowField2023}
{\sc \au{Huang, S.-Y.}, \au{Zhou, J.}, \au{Liu, S.-J.}, \au{Peng, H.-Y.},
  \au{Yuan, X.-Q.} \& \au{Zhang, H.-L.}} \yr{2023{\natexlab{{\em a\/}}}}
  \at{Propagation and flow field analysis of wall-detached continuous rotating
  detonation wave in a hollow combustor}.  \jt{Combust. Flame}  \bvol{248},
  \pg{112550}.

\bibitem[Huang {\em et~al.\/}(2023{\natexlab{{\em b\/}}})Huang, Zhou, Liu, Liu,
  Peng, Zhang \& Yuan]{huangAnalysisRadialStructure2023}
{\sc \au{Huang, S.-Y.}, \au{Zhou, J.}, \au{Liu, W.-D.}, \au{Liu, S.-J.},
  \au{Peng, H.-Y.}, \au{Zhang, H.-L.} \& \au{Yuan, X.-Q.}}
  \yr{2023{\natexlab{{\em b\/}}}}  \at{Analysis on the radial structure of
  rotating detonation wave in a hollow combustor}.  \jt{Fuel}  \bvol{348},
  \pg{128581}.

\bibitem[Katta {\em et~al.\/}(2019)Katta, Cho, Hoke, Codoni, Schauer \&
  Roquemore]{kattaEffectIncreasingChannel2019}
{\sc \au{Katta, V.~R.}, \au{Cho, K.~Y.}, \au{Hoke, J.~L.}, \au{Codoni, J.~R.},
  \au{Schauer, F.~R.} \& \au{Roquemore, W.~M.}} \yr{2019}  \at{Effect of
  increasing channel width on the structure of rotating detonation wave}.
  \jt{Proc. Combust. Inst.}  \bvol{37}~(3),  \pg{3575--3583}.

\bibitem[Liu {\em et~al.\/}(2026)Liu, Wang, Zhao, Weng \&
  Zhao]{liuRolesAcousticImpedance2026}
{\sc \au{Liu, Q.}, \au{Wang, F.}, \au{Zhao, F.}, \au{Weng, C.} \& \au{Zhao,
  Q.}} \yr{2026}  \at{Roles of acoustic impedance and acoustic velocity ratios
  on semiconfined detonation propagation}.  \jt{AIAA J.}  \bvol{64}~(4),
  \pg{2024--2039}.

\bibitem[Ma {\em et~al.\/}(2025)Ma, Liu, Li \&
  Ning]{maNumericalStudyPropagation2025}
{\sc \au{Ma, T.}, \au{Liu, X.}, \au{Li, J.} \& \au{Ning, J.}} \yr{2025}
  \at{Numerical study on propagation of detonation waves in a non-uniform
  mixture with transverse temperature discontinuities}.  \jt{J. Fluid Mech.}
  \bvol{1022},  \pg{A6}.

\bibitem[Nassini {\em et~al.\/}(2023)Nassini, Andreini \&
  Bohon]{nassiniCharacterizationRefillRegion2023}
{\sc \au{Nassini, P.~C.}, \au{Andreini, A.} \& \au{Bohon, M.~D.}} \yr{2023}
  \at{Characterization of refill region and mixing state immediately ahead of a
  hydrogen-air rotating detonation using {LES}}.  \jt{Combust. Flame}
  \bvol{258},  \pg{113050}.

\bibitem[Nejaamtheen \&
  Choi(2026)]{nejaamtheenThreedimensionalMixingDetonation2026}
{\sc \au{Nejaamtheen, M.} \& \au{Choi, J.-Y.}} \yr{2026}  \at{Three-dimensional
  mixing and detonation stability in disk-shaped rotating detonation rocket
  engines: {A} comparative study of discrete and slit injectors}.  \jt{Combust.
  Flame}  \bvol{288},  \pg{114931}.

\bibitem[Nejaamtheen {\em et~al.\/}(2021)Nejaamtheen, Kim, Pavalavanni, Ryu \&
  Choi]{nejaamtheenEffectsDimensionlessRadius2021}
{\sc \au{Nejaamtheen, M.~N.}, \au{Kim, T.~Y.}, \au{Pavalavanni, P.~K.},
  \au{Ryu, J.} \& \au{Choi, J.-Y.}} \yr{2021}  \at{Effects of the dimensionless
  radius of an annulus on the detonation propagation characteristics in
  circular and non-circular rotating detonation engines}.  \jt{Shock Waves}
  \bvol{31}~(7),  \pg{703--715}.

\bibitem[Petersen {\em et~al.\/}(2003)Petersen, Kalitan \&
  Rickard]{petersenCalibrationChemicalKinetics2003}
{\sc \au{Petersen, E.~L.}, \au{Kalitan, D.~M.} \& \au{Rickard, M. J.~A.}}
  \yr{2003} Calibration and chemical kinetics modeling of an {OH}
  chemiluminescence diagnostic.  \bt{In {\em 39th {AIAA}/{ASME}/{SAE}/{ASEE}
  {Joint} {Propulsion} {Conference} and {Exhibit}\/}}.  \publ{Huntsville,
  Alabama: American Institute of Aeronautics and Astronautics}, {AIAA} Paper
  2003-4493.

\bibitem[Raj \& Meadows(2024)]{rajInfluenceFuelInhomogeneity2024}
{\sc \au{Raj, P.} \& \au{Meadows, J.}} \yr{2024}  \at{Influence of fuel
  inhomogeneity on detonation wave propagation in a rotating detonation
  combustor}.  \jt{Shock Waves}  \bvol{34}~(5),  \pg{429--449}.

\bibitem[Raman {\em et~al.\/}(2023)Raman, Prakash \&
  Gamba]{ramanNonidealitiesRotatingDetonation2023}
{\sc \au{Raman, V.}, \au{Prakash, S.} \& \au{Gamba, M.}} \yr{2023}
  \at{Nonidealities in rotating detonation engines}.  \jt{Annu. Rev. Fluid
  Mech.}  \bvol{55}~(1),  \pg{639--674}.

\bibitem[Rankin {\em et~al.\/}(2019)Rankin, Codoni, Cho, Hoke \&
  Schauer]{rankinInvestigationStructureDetonation2019}
{\sc \au{Rankin, B.~A.}, \au{Codoni, J.~R.}, \au{Cho, K.~Y.}, \au{Hoke, J.~L.}
  \& \au{Schauer, F.~R.}} \yr{2019}  \at{Investigation of the structure of
  detonation waves in a non-premixed hydrogen–air rotating detonation engine
  using mid-infrared imaging}.  \jt{Proc. Combust. Inst.}  \bvol{37}~(3),
  \pg{3479--3486}.

\bibitem[Rankin {\em et~al.\/}(2017)Rankin, Fotia, Naples, Stevens, Hoke,
  Kaemming, Theuerkauf \& Schauer]{rankinOverviewPerformanceApplication2017}
{\sc \au{Rankin, B.~A.}, \au{Fotia, M.~L.}, \au{Naples, A.~G.}, \au{Stevens,
  C.~A.}, \au{Hoke, J.~L.}, \au{Kaemming, T.~A.}, \au{Theuerkauf, S.~W.} \&
  \au{Schauer, F.~R.}} \yr{2017}  \at{Overview of performance, application, and
  analysis of rotating detonation engine technologies}.  \jt{J. Propul. Power}
  \bvol{33}~(1),  \pg{131--143}.

\bibitem[Rankin {\em et~al.\/}(2016)Rankin, Fugger, Richardson, Cho, Hoke,
  Caswell, Gord \& Schauer]{rankinEvaluationMixingProcesses2016}
{\sc \au{Rankin, B.~A.}, \au{Fugger, C.~A.}, \au{Richardson, D.~R.}, \au{Cho,
  K.~Y.}, \au{Hoke, J.~L.}, \au{Caswell, A.~W.}, \au{Gord, J.~R.} \&
  \au{Schauer, F.~R.}} \yr{2016} Evaluation of mixing processes in a
  non-premixed rotating detonation engine using acetone {PLIF} imaging.  \bt{In
  {\em 54th {AIAA} {Aerospace} {Sciences} {Meeting}\/}}.  \publ{San Diego,
  California, USA: American Institute of Aeronautics and Astronautics}, {AIAA}
  Paper 2016-1198.

\bibitem[Sato {\em et~al.\/}(2021)Sato, Chacon, White, Raman \&
  Gamba]{satoMixingDetonationStructure2021a}
{\sc \au{Sato, T.}, \au{Chacon, F.}, \au{White, L.}, \au{Raman, V.} \&
  \au{Gamba, M.}} \yr{2021}  \at{Mixing and detonation structure in a rotating
  detonation engine with an axial air inlet}.  \jt{Proc. Combust. Inst.}
  \bvol{38}~(3),  \pg{3769--3776}.

\bibitem[Schwer \& Kailasanath(2011)]{schwerNumericalInvestigationPhysics2011}
{\sc \au{Schwer, D.} \& \au{Kailasanath, K.}} \yr{2011}  \at{Numerical
  investigation of the physics of rotating-detonation-engines}.  \jt{Proc.
  Combust. Inst.}  \bvol{33}~(2),  \pg{2195--2202}.

\bibitem[Shi {\em et~al.\/}(2025)Shi, Dai, Zhao, Wen, Wang, Ma \&
  Mou]{shiCalculationModelWall2025}
{\sc \au{Shi, Y.}, \au{Dai, W.}, \au{Zhao, P.}, \au{Wen, H.}, \au{Wang, B.},
  \au{Ma, G.} \& \au{Mou, H.}} \yr{2025}  \at{Calculation model of wall heat
  flux in rotating detonation combustor and its experimental validation}.
  \jt{Appl. Therm. Eng.}  \bvol{280},  \pg{128100}.

\bibitem[Smith {\em et~al.\/}(2002)Smith, Luque, Park, Jeffries \&
  Crosley]{smithLowPressureFlame2002}
{\sc \au{Smith, G.~P.}, \au{Luque, J.}, \au{Park, C.}, \au{Jeffries, J.~B.} \&
  \au{Crosley, D.~R.}} \yr{2002}  \at{Low pressure flame determinations of rate
  constants for {OH}({A}) and {CH}({A}) chemiluminescence}.  \jt{Combust.
  Flame}  \bvol{131}~(1-2),  \pg{59--69}.

\bibitem[Strempfl {\em et~al.\/}(2024)Strempfl, Dounia, Laera \&
  Poinsot]{strempflEffectsMixingAssumptions2024}
{\sc \au{Strempfl, P.}, \au{Dounia, O.}, \au{Laera, D.} \& \au{Poinsot, T.}}
  \yr{2024}  \at{Effects of mixing assumptions and models for {LES} of
  hydrogen-fueled rotating detonation engines}.  \jt{Int. J. Hydrogen Energy}
  \bvol{62},  \pg{1--16}.

\bibitem[Sun {\em et~al.\/}(2019)Sun, Zhou, Liu \&
  Lin]{sunInteractionRotatingDetonation2019}
{\sc \au{Sun, J.}, \au{Zhou, J.}, \au{Liu, S.} \& \au{Lin, Z.}} \yr{2019}
  \at{Interaction between rotating detonation wave propagation and reactant
  mixing}.  \jt{Acta Astronaut.}  \bvol{164},  \pg{197--203}.

\bibitem[Tamura {\em et~al.\/}(1998)Tamura, Berg, Harrington, Luque, Jeffries,
  Smith \& Crosley]{tamuraCollisionalQuenchingCHA1998}
{\sc \au{Tamura, M.}, \au{Berg, P.~A.}, \au{Harrington, J.~E.}, \au{Luque, J.},
  \au{Jeffries, J.~B.}, \au{Smith, G.~P.} \& \au{Crosley, D.~R.}} \yr{1998}
  \at{Collisional quenching of {CH}({A}), {OH}({A}), and {NO}({A}) in low
  pressure hydrocarbon flames}.  \jt{Combust. Flame}  \bvol{114}~(3-4),
  \pg{502--514}.

\bibitem[Tang {\em et~al.\/}(2015)Tang, Wang \&
  Shao]{tangThreedimensionalNumericalInvestigations2015}
{\sc \au{Tang, X.-M.}, \au{Wang, J.-P.} \& \au{Shao, Y.-T.}} \yr{2015}
  \at{Three-dimensional numerical investigations of the rotating detonation
  engine with a hollow combustor}.  \jt{Combust. Flame}  \bvol{162}~(4),
  \pg{997--1008}.

\bibitem[Uemura {\em et~al.\/}(2013)Uemura, Hayashi, Asahara, Tsuboi \&
  Yamada]{uemuraTransverseWaveGeneration2013}
{\sc \au{Uemura, Y.}, \au{Hayashi, A.~K.}, \au{Asahara, M.}, \au{Tsuboi, N.} \&
  \au{Yamada, E.}} \yr{2013}  \at{Transverse wave generation mechanism in
  rotating detonation}.  \jt{Proc. Combust. Inst.}  \bvol{34}~(2),
  \pg{1981--1989}.

\bibitem[Ullman \& Raman(2024)]{ullmanInfluenceInjectorConfiguration2024}
{\sc \au{Ullman, M.} \& \au{Raman, V.}} \yr{2024}  \at{Influence of injector
  configuration on secondary wave formation and propagation in a rotating
  detonation combustor}.  \jt{Appl. Energy Combust. Sci.}  \bvol{20},
  \pg{100289}.

\bibitem[Valencia {\em et~al.\/}(2024)Valencia, Mendiburu, Bravo, Khare \&
  Celis]{valenciaFlowfieldAnalysisPerformance2024}
{\sc \au{Valencia, S.}, \au{Mendiburu, A.}, \au{Bravo, L.}, \au{Khare, P.} \&
  \au{Celis, C.}} \yr{2024}  \at{Flow-field analysis and performance assessment
  of rotating detonation engines under different number of discrete inlet
  nozzles}.  \jt{Appl. Energy Combust. Sci.}  \bvol{20},  \pg{100296}.

\bibitem[Vignat {\em et~al.\/}(2024)Vignat, Brouzet, Bonanni \&
  Ihme]{vignatAnalysisWeakSecondary2024}
{\sc \au{Vignat, G.}, \au{Brouzet, D.}, \au{Bonanni, M.} \& \au{Ihme, M.}}
  \yr{2024}  \at{Analysis of weak secondary waves in a rotating detonation
  engine using large-eddy simulation and wavenumber-domain filtering}.
  \jt{Combust. Flame}  \bvol{263},  \pg{113387}.

\bibitem[Wahba {\em et~al.\/}(2026)Wahba, Mi, Kiyanda \&
  Higgins]{wahbaDetonationPropagationWeakly2026a}
{\sc \au{Wahba, Y.~K.}, \au{Mi, X.~C.}, \au{Kiyanda, C.~B.} \& \au{Higgins,
  A.~J.}} \yr{2026}  \at{Detonation propagation in weakly confined gases}.
  \jt{Shock Waves}  \bvol{36}~(3),  \pg{22}.

\bibitem[Wang {\em et~al.\/}(2024{\natexlab{{\em a\/}}})Wang, Webb,
  Athmanathan, Slipchenko, Kearney, Perkins, Roy, Fugger \&
  Meyer]{wang500kHzOHPLIF2024}
{\sc \au{Wang, R.~B.}, \au{Webb, A.~M.}, \au{Athmanathan, V.}, \au{Slipchenko,
  M.~N.}, \au{Kearney, S.~P.}, \au{Perkins, H.~D.}, \au{Roy, S.}, \au{Fugger,
  C.~A.} \& \au{Meyer, T.~R.}} \yr{2024{\natexlab{{\em a\/}}}}  \at{500-{kHz}
  {OH} {PLIF} and {OH}* chemiluminescence imaging of deflagration and rotating
  detonation in {CH4}-{O2} and {H2}-air mixtures}.  \jt{Proc. Combust. Inst.}
  \bvol{40}~(1-4),  \pg{105770}.

\bibitem[Wang {\em et~al.\/}(2024{\natexlab{{\em b\/}}})Wang, Webb, Camacho,
  Slipchenko, Meyer, Athmanathan, Braun, Fugger, Roy \&
  Perkins]{wangHighspeedChemiluminscenceOHPLIF2024}
{\sc \au{Wang, R.~B.}, \au{Webb, A.~M.}, \au{Camacho, R.~D.}, \au{Slipchenko,
  M.~N.}, \au{Meyer, T.~R.}, \au{Athmanathan, V.}, \au{Braun, J.}, \au{Fugger,
  C.~A.}, \au{Roy, S.} \& \au{Perkins, H.~D.}} \yr{2024{\natexlab{{\em b\/}}}}
  Characterization of reactant refill and detonation wave dynamics in a natural
  gas/{GOx} rotating detonation engine using simultaneous high repetition-rate
  {OH}-{PLIF} and chemiluminescence.  \bt{In {\em {AIAA} {SCITECH} 2024
  {Forum}\/}}.  \publ{Orlando, FL: American Institute of Aeronautics and
  Astronautics}, {AIAA} Paper 2024-1032.

\bibitem[Wang {\em et~al.\/}(2023)Wang, Huang, Deiterding, Chen \&
  Chen]{wangNumericalStudiesDetonation2023}
{\sc \au{Wang, Y.}, \au{Huang, C.}, \au{Deiterding, R.}, \au{Chen, H.} \&
  \au{Chen, Z.}} \yr{2023}  \at{Numerical studies on detonation propagation in
  inhomogeneous mixtures with periodic reactant concentration gradient}.
  \jt{J. Fluid Mech.}  \bvol{955},  \pg{A23}.

\bibitem[Wen {\em et~al.\/}(2026)Wen, Luo, Xu \&
  Wang]{wenJANCCosteffectiveDifferentiable2026}
{\sc \au{Wen, H.}, \au{Luo, F.}, \au{Xu, S.} \& \au{Wang, B.}} \yr{2026}
  \at{{JANC}: {A} cost-effective, differentiable compressible reacting flow
  solver featured with {JAX}-based adaptive mesh refinement}.  \jt{Comput.
  Phys. Commun.}  \bvol{319},  \pg{109915}.

\bibitem[Wen {\em et~al.\/}(2019)Wen, Xie \&
  Wang]{wenPropagationBehaviorsRotating2019}
{\sc \au{Wen, H.}, \au{Xie, Q.} \& \au{Wang, B.}} \yr{2019}  \at{Propagation
  behaviors of rotating detonation in an obround combustor}.  \jt{Combust.
  Flame}  \bvol{210},  \pg{389--398}.

\bibitem[Wolański(2013)]{wolanskiDetonativePropulsion2013}
{\sc \au{Wolański, P.}} \yr{2013}  \at{Detonative propulsion}.  \jt{Proc.
  Combust. Inst.}  \bvol{34}~(1),  \pg{125--158}.

\bibitem[Yokoo {\em et~al.\/}(2021)]{yokooExperimentalStudyInternal2021}
{\sc \au{Yokoo, R.} \& \au{others}} \yr{2021}  \at{Experimental study of
  internal flow structures in cylindrical rotating detonation engines}.
  \jt{Proc. Combust. Inst.}  \bvol{38}~(3),  \pg{3759--3768}.

\bibitem[Zhao {\em et~al.\/}(2018)Zhao, Buttsworth \&
  Choudhury]{zhaoExperimentalNumericalStudy2018}
{\sc \au{Zhao, M.}, \au{Buttsworth, D.} \& \au{Choudhury, R.}} \yr{2018}
  \at{Experimental and numerical study of {OH}* chemiluminescence in hydrogen
  diffusion flames}.  \jt{Combust. Flame}  \bvol{197},  \pg{369--377}.

\bibitem[Zhou \& Wang(2013)]{zhouNumericalInvestigationShock2013}
{\sc \au{Zhou, R.} \& \au{Wang, J.-P.}} \yr{2013}  \at{Numerical investigation
  of shock wave reflections near the head ends of rotating detonation engines}.
   \jt{Shock Waves}  \bvol{23}~(5),  \pg{461--472}.

\end{thebibliography}
\end{document}